\documentclass[aps,prd,superscriptaddress,nofootinbib,11pt]{revtex4}
\usepackage[english]{babel}
\usepackage[utf8]{inputenc}
\usepackage{graphicx}   % need for figures
\usepackage{slashed}
\usepackage{epstopdf}
\usepackage{verbatim}   % useful for program listings
\usepackage{color}      % use if color is used in text
\usepackage{subfigure}  % use for side-by-side figures
\usepackage{multirow}
\usepackage{hyperref}   % use for hypertext links, including those to external documents and URLs
\usepackage{float}
\usepackage{epsfig,rotating}
\usepackage{amsmath,amssymb}
\usepackage{dsfont}
\restylefloat{table}
\numberwithin{equation}{section}

\newcommand{\vx}{\vec{x}}
\newcommand{\vp}{\vec{p}}
\newcommand{\vq}{\vec{q}}
\newcommand{\vk}{\vec{k}}

\newcommand{\be}{\begin{equation}}
\newcommand{\ee}{\end{equation}}
\newcommand{\bea}{\begin{eqnarray}}
\newcommand{\eea}{\end{eqnarray}}

\newcommand{\ket}[1]{|#1\rangle}
\newcommand{\bra}[1]{\langle#1|}

\begin{document}
%\tableofcontents
\title{Quantum decay   in renormalizable field theories:\\ quasiparticle formation, Zeno  and anti-Zeno effects.}
\author{Daniel Boyanovsky}
\email{boyan@pitt.edu} \affiliation{Department of Physics and
Astronomy, University of Pittsburgh, Pittsburgh, PA 15260}
\date{\today}

\begin{abstract}
In a renormalizable theory the survival probability of an unstable quantum state features  divergences as a consequence of the rapid growth of the density of states with energy. Introducing a high energy cutoff $\Lambda$, the transient dynamics during a time scale $\simeq 1/\Lambda$ describes the renormalization of the bare into a ``quasiparticle'' state which decays on longer time scales. During this early transient the decay law features   Zeno behavior $e^{-(t/t_Z)^2}$ with the Zeno time-scale $t_Z \propto 1/\Lambda^{3/2}$. We introduce a dynamical renormalization framework that allows to separate consistently the dynamics of formation of the quasiparticle state and its decay on longer time scales by introducing a renormalization time scale along with alternative ``schemes''. The survival probability obeys a renormalization group equation with respect to this scale. We find a transient \emph{suppression} of the decay law for large Lorentz factor as a consequence of the narrowing of the phase space for decay different from the usual time dilation. In presence of higher mass thresholds, the energy uncertainty associated with transient dynamics leads to an \emph{anti Zeno enhancement} with a transient acceleration of the decay into \emph{heavier particles}. There remains memory of the transient effects in the survival probability even at long time. We discuss possible consequences of these effects in cosmology.
\end{abstract}


\maketitle

\section{Introduction}
The decay of unstable quantum states is of fundamental interdisciplinary importance  in  particle physics, condensed matter and quantum optics. Most of the particles in the standard model decay (typically into various channels) and their decay properties provide observational hallmarks. Particle decay   also plays an important role in various processes in early Universe cosmology\cite{kolb,weinberg,decaybaryo,decaybaryo2,wimpdmlepto,leptodecay,roulet}.   The fundamental description of the time evolution of unstable states in quantum field theory has been the focus of early studies\cite{khalfin,peres,fonda,maiani} revealing surprising phenomena such as the Zeno effect resulting in an early time delay of decay\cite{misra,naka,review,zenopasca} as well as long time non-exponential decay\cite{fonda,khalfin,maiani}. These  important aspects of the early and late time evolution of unstable states in quantum field theory continue to be studied at a deeper level\cite{naka,faccipasca,facchi,cosa,yukacosa,supercosa,anasto,urbano,giunti,ramirez}.

In quantum field theory there are two main approaches to studying the decay of a particle: a) within the S-matrix framework the decay rate is obtained from the transition probability of a single particle ``in'' state prepared in the infinite past, to the decay product ``out'' state in the infinite future. The probability features the square of an energy conserving delta function that is interpreted as overall energy conservation multiplied by the total time elapsed, the total probability divided by this total time is identified with the decay rate. b) An alternative framework is to obtain the Dyson-summed propagator including self-energy corrections for the decaying particle, the imaginary part of the self-energy on the ``mass shell'' of the particle is identified with the decay rate. The full propagator is approximated by a Breit-Wigner distribution, which describes the decaying state as a pole in the second Riemann sheet in the complex frequency plane. The inverse Fourier transform of the full propagator (in the Breit-Wigner representation) yields the forward time evolution   of this state, which in the long time limit is dominated by the complex pole near the real axis.
While these two (equivalent) methods are ubiquitous in phenomenology of particle physics and generically in the description of unstable states from S-matrix theory, they are not suitable within the cosmological framework where the time evolution of the gravitational background prevents energy conservation as a consequence of the absence of a global time-like Killing vector\cite{parker,birrell,fullbook,mukhabook,parkerbook}. For example, the local energy of a particle of mass $M$ and comoving momentum $k$ as measured by a comoving observer is $E_k(t)= \sqrt{\frac{k^2}{a^2(t)}+M^2}$ where $a(t)$ is the time dependent scale factor that defines the expanding spatially flat Friedmann-Robertson-Walker cosmology.  Furthermore taking the infinite time limit in this case   is obviously not justified as the background metric evolves in time,  thus excluding a time Fourier transform of propagators and self-energies. Instead, the time evolution of decaying states must be studied directly in real time. Early studies focused on describing the dynamics of decaying states in inflationary\cite{boydecay,mosca} and standard cosmology\cite{spangdecay}, and more recent studies have addressed the time evolution of quantum unstable states in radiation and matter dominated cosmologies\cite{vilja,herring}.

\vspace{1mm}

\textbf{Motivations and goals:}
The studies of particle decay in cosmology reported in references\cite{vilja,herring} focused on super renormalizable theories. In reference\cite{vilja} the authors analyzed the decay of a bosonic particle into two other bosonic particles by obtaining the solutions of the free field Heisenberg equations in the background of a radiation or stiff matter dominated cosmology. Reference\cite{herring} studied the decay in a similar model introducing a physically motivated adiabatic approximation combined with a non-perturbative formulation of the time evolution of states. This method is a field theoretical extension of the Weisskopf-Wigner theory of atomic linewidth\cite{ww} ubiquitous in quantum optics\cite{book} which has been generalized to study time evolution in particle physics\cite{boyww} and within cosmological settings\cite{boycosww}. As we discuss in this article, the  time evolution of unstable states in \emph{renormalizable} quantum field theories features novel aspects that are not present in super-renormalizable theories. The main difference is traced to the ultraviolet divergences associated with the dressing of the bare states by many body correlations, which lead to the formation of a \emph{quasiparticle} state in a renormalizable theory.

Our main motivation  is to ultimately provide a systematic framework to describe decay processes (and   their inverse) in an expanding cosmology. Towards this goal, in this article we study particle decay in Minkowski space-time focusing on two main aspects: i) to develop a non-perturbative formulation of the time evolution during the decay process that includes consistently the \emph{renormalization} of bare states,  namely the formation of the ``dressed'' quasi particle states during the transient dynamics, and their  decay on longer time scales. ii) to study transient phenomena   that contribute  to the decay process and affects the survival probability, but that is \emph{not} reliably described by the S-matrix approach.

\vspace{1mm}

\textbf{Summary of results:} The survival probability at time $t$ of an initial quantum state with momentum $\vk$ prepared at $t=0$    can be written as $P_k(t) = e^{-\alpha_k(t)}\,P_k(0)$. In a renormalizable theory the decay function $\alpha_k(t)$ features ultraviolet divergences, which are manifest during a time scale $\simeq 1/\Lambda$ with $\Lambda$ a high energy cutoff. This early transient describes the   renormalization of the bare state into a state ``dressed'' by many body correlations, namely the build-up of a \emph{quasiparticle} state   which decays on a longer time scale. During this transient stage,  the decay function features a Zeno-like behavior $\alpha_k(t) =(t/t_Z)^2$\cite{naka,review,zenopasca,faccipasca,facchi} with $t_Z \propto 1/\Lambda^{3/2}$. We introduce a dynamical renormalization framework along with alternative minimal and ``on-shell'' schemes  that allow a consistent separation between the dynamics of formation and decay of the quasiparticle state. This is achieved by  introducing an arbitrary renormalization time scale $t_b$ intermediate between the formation and decay of the quasiparticle. The survival probability is \emph{independent} of this scale and consequently obeys a renormalization group equation with respect to $t_b$. The transient dynamics   reveals a \emph{delay} in decay   as a consequence of the narrowing of the phase space for large Lorentz factor and is \emph{independent} of the usual time dilation.   In the presence of several channels arising from coupling to higher mass particles, the energy uncertainty associated with transient dynamics allows to probe the higher mass thresholds. This results in an \emph{anti-Zeno} effect\cite{antizenokuri,lewantizeno} and a concomitant \emph{acceleration} of decay into heavier particles during a transient period which scales $\simeq 1/\gamma_k$ with $\gamma_k$ the Lorentz factor. At very long time scales, well after the transient phenomena has subsided, the linear time behavior described by S-matrix theory and exponential decay emerges but with important corrections from the  transient behavior. Asymptotically, the \emph{memory} of the  transient dynamics is imprinted in the decay function as an overall constant, which, however, yields a  further suppression of  the survival probability beyond the usual decay with a constant decay rate from the S-matrix approach.

\section{Renormalizable and super-renormalizable models.}

In order to highlight the differences and to compare the decay dynamics between the renormalizable and super-renormalizable cases we consider the following two models.

\vspace{2mm}

\textbf{Renormalizable:} as a renormalizable theory we study a real scalar (Higgs) field  $\Phi$ of mass $M$  Yukawa coupled to fermionic fields $\Psi_f$, with Lagrangian density
\be \mathcal{L} = \frac{1}{2} \partial_\mu \Phi \partial^\mu \Phi -\frac{1}{2} M^2 \,\Phi^2 + \sum_{f} \Big\{ \overline{\Psi}_f\big(i\gamma^\mu \partial_\mu - m_f\big)\Psi_f - Y_f\,\Phi \,\overline{\Psi}_f\,\Psi_f\Big\} \,, \label{Lyuk}\ee where the index $f$ refers to different ``flavors'' with a hierarchy of masses $m_1< m_2 \cdots$ and $Y_f$ are the respective Yukawa couplings.  We consider Dirac fields, the extension to Majorana fields is straightforward yielding similar results.

\vspace{2mm}

\textbf{Super renormalizable} as an example of a super renormalizable theory we consider the same real scalar (Higgs) field  $\Phi$  with a cubic coupling to other real scalar fields $\chi_f$, with cubic couplings $\lambda_f$,  namely

\be \mathcal{L} = \frac{1}{2} \partial_\mu \Phi \partial^\mu \Phi -\frac{1}{2} M^2 \,\Phi^2 + \sum_{f} \Big\{ \frac{1}{2} \partial_\mu \chi_f \partial^\mu \chi_f -\frac{1}{2} m^2_f \,\chi^2_f   - \lambda_f\,\Phi \,\chi^2_f \Big\} \,, \label{Lcub}\ee again assuming a hierarchy of masses. The main reason for considering coupling to several fields with increasing masses is to include higher mass thresholds in order to analyze the enhancement of decay as a consequence of energy uncertainties during the transient evolution, described in section (\ref{sec:antizeno}).

Free field quantization in the discrete momentum representation in  a finite volume $V$  proceeds as usual. For a generic real scalar field $\varphi(x)$  of mass m:
\be \varphi(x) =  \sum_{\vk}\,\frac{1}{\sqrt{2VE_k}} \Big[a_{\vk} \,e^{-iE_k t}\,e^{i\vk\cdot\vx} + a^\dagger_{\vk} \,e^{iE_k t}\,e^{-i\vk\cdot\vx}   \Big] ~~;~~ E_k = \sqrt{k^2+m^2} \,, \label{boseexp}\ee and for a generic (Dirac) fermi field $(\psi)$ of mass $m$

\be \psi(x) =  \,\sum_{\vk,s}\,\frac{1}{\sqrt{2VE_k}} \Big[b_{\vk,s}\, U_{\vk,s}  \,e^{-iE_k t}\,e^{i\vk\cdot\vx} + d^\dagger_{\vk,s}\,V_{\vk,s}  \,e^{iE_k t}\,e^{-i\vk\cdot\vx}   \Big]\,, \label{fermiexp}\ee where the spinors $U,V$ are normalized as
\be U^\dagger_{\vk,s} U_{\vk,s'} = 2 \,E_k \,\delta_{s,s'}=  V^\dagger_{\vk,s} V_{\vk,s'}\,,  \label{normas}\ee  and obey

\be \sum_{s}U_{\vk,s,a}\overline{U}_{\vk,s,b} = (\gamma^\mu K_\mu + m)_{ab} ~~;~~ \sum_{s}V_{\vk,s,a}\overline{V}_{\vk,s,b} = (\gamma^\mu K_\mu - m)_{ab} ~~;~~ K^\mu= (E_k, \vk) \,. \label{projec} \ee We use the standard Dirac representation for the $\gamma$ matrices, along with the metric $g_{\mu\nu} = \textrm{diag}(1,-1,-1,-1)$.

\section{Survival probability}
\subsection{ Perturbation theory.}
Consider an initial state described by a single particle state of the Higgs field $\Phi$ with momentum $\vk$ namely $\ket{1^{\Phi}_{\vk}}$, the persistence  or survival amplitude at time $t$ is given by
\be  {A}_k(t) = \bra{1^{\Phi}_{\vk}}e^{-iHt}\ket{1^{\Phi}_{\vk}}\,, \label{amp}\ee where  $H=H_0+H_i$ is the total Hamiltonian, with $H_0,H_i$ the free field and interaction Hamiltonians respectively. Passing to the interaction picture with $e^{-iHt} = e^{-iH_0t}\,U_I(t)$ where   $U_I(t)$ is the unitary time evolution operator in the interaction picture,
\be U_I(t) = 1 - i\int^t_0 H_I(t')\,dt' - \int^t_0 dt_1\,\int^{t_1}_0 dt_2 H_I(t_1)\,H_I(t_2) + \cdots ~~;~~ H_I(t) = e^{iH_0t} \,H_i \, e^{-iH_0t}\,. \label{UI}\ee  Up to second order in the interaction   the survival probability is obtained by introducing a complete set of intermediate eigenstates of $H_0$, namely  $H_0 \ket{n} = \varepsilon_n \ket{n}$,  we find
\be  {P}_k(t) = | {A}_k(t)|^2 = 1- 2 \sum_{n} \frac{|\bra{1^{\Phi}_{\vk}}H_i\ket{n}|^2}{(\omega_k-\varepsilon_n)^2}\, \Big[1-\cos\big[\big( \omega_k-\varepsilon_n\big)t\big]\Big]+ \cdots ~~;~~\omega_k = \sqrt{k^2+M^2} \label{Poft}\ee  For the models described above there are several intermediate states that contribute:   a disconnected (vacuum) contribution with $\ket{n}$ being a four particles state in which the initial state evolves without any interaction, and a connected contribution in which the initial state decays into two scalars  or a fermion-anti-fermion pair respectively.

The disconnected diagrams yield an infinite phase that multiplies the state $\ket{1^{\Phi}_{\vk}}$, this phase is irrelevant to the survival probability and decay law and is not related to the main decay process, therefore it will not be considered further.

Only the connected contribution is relevant for the survival probability (see below).
We introduce the spectral density
\be \rho(k_0,k) = \sum_{\kappa}|\bra{1^{\Phi}_{\vk}}H_i\ket{\kappa}|^2\,\delta(k_0 -\varepsilon_\kappa) \label{specdens}\ee to write the survival probability (\ref{Poft}) as
\be    {P}_k(t) = 1- 2\int^{\infty}_{-\infty} dk_0 \, \rho(k_0,k) \,\frac{\Big[1-\cos\big[\big( k_0-\omega_k\big)t\big]\Big]}{(k_0-\omega_k)^2}\,. \label{Poftrho}\ee The early time behavior of the survival probability is given by
\be P_k(t) \simeq 1- t^2/t^2_Z ~~;~~ 1/t^2_Z = \int^{\infty}_{-\infty} dk_0 \, \rho(k_0,k)\,,\label{tzeno}\ee where $t_Z$ is identified as the Zeno time\cite{misra,fonda,maiani,review,naka,zenopasca,faccipasca,facchi} which from the definition (\ref{specdens}) can also be written as
\be \frac{1}{t^2_Z} = |\bra{1^{\Phi}_{\vk}}H^2_i\ket{1^{\Phi}_{\vk}}| \,.\label{H2tz}\ee
 In general in a quantum field theory the integral in (\ref{tzeno}) diverges, cutting the $k_0$ integral off at a scale $\Lambda \gg \omega_k$, the early time approximation (\ref{tzeno}) is valid for $t \lesssim 1/\Lambda$, which in general will be unobservable. However, we argue below that this time scale is very important in the renormalizable case as it determines the renormalization of the bare state and the formation of a \emph{quasiparticle} state.

Up to second order in the coupling, the intermediate states for the Yukawa theory (renormalizable) correspond to a fermion antifermion pair $\ket{n} \equiv \ket{1_{\vp,s},\overline{1}_{\vq,s'}}$  whereas for the super renormalizable model the intermediate state correspond to a pair of bosonic $\chi$ particles $\ket{n} = \ket{1_{\vp},{1}_{\vq}}$,   in both cases $\vq = \vk-\vp$.  We now consider only one species of either bosonic or fermionic fields and  find for the super renormalizable (S) and renormalizable (R) cases the following spectral densities corresponding to one-loop self-energy (connected) contributions, (details are given in the appendix)
\be \rho_S(k_0,k) = \frac{\lambda^2}{16\pi^2 \,\omega_k}\, \sqrt{1-\frac{4m^2}{k^2_0-k^2}}\,~~\Theta(k_0)\,\Theta(k^2_0-k^2-4m^2)   \,, \label{rhoS}\ee

\be \rho_R(k_0,k) = \frac{Y^2}{16\pi^2 \,\omega_k}\, (k^2_0-k^2) \Bigg[1-\frac{4m^2}{k^2_0-k^2}\Bigg]^{3/2}\,~~\Theta(k_0)\,\Theta(k^2_0-k^2-4m^2)~~;~~\omega_k = \sqrt{k^2+M^2} \,, \label{rhoR}\ee where $M$ is the mass of the scalar field $\Phi$ associated with the decaying particle, and $m$ that of the decay product. The theta functions in the above expressions describe the two particle threshold. Including the other ``flavors'' with masses $m_f$ there is a new threshold for the production of each pair corresponding to different decay channels. The total spectral density is the sum of all the individual spectral densities.

Using the following identity in the sense of distribution

\be \frac{\Big[1-\cos\big[\big( k_0-\omega_k\big)t\big]\Big]}{(k_0-\omega_k)^2} ~~{}_{\overrightarrow{t\rightarrow \infty}}  ~~ \pi\, t \, \delta(k_0-\omega_k) + \mathcal{P}\,\Bigg( \frac{1}{(k_0-\omega_k)^2}\Bigg)\,, \label{iden}\ee where $\mathcal{P}$ stands for the principal part, we find that consistently up to second order in the interaction, the formal long time limit of the  survival probability (\ref{Poft}) becomes
\be  {P}(t) \simeq  {Z}^2\,\Big[1- \Gamma_k t  \Big] +\cdots \label{asyPoft}\ee
where $ {Z}$ is the wavefunction renormalization constant, given by
\be  {Z} = 1- \int^{\infty}_{-\infty}\mathcal{P} \Bigg(\frac{\rho(k_0,k)}{(k_0-\omega_k)^2} \Bigg) \,dk_0 = 1- {\sum_{n}}^{\,'}\frac{|\bra{1_{\vk}}H_i\ket{n}|^2}{(\omega_k-\varepsilon_n)^2} < 1\,, \label{waf}\ee where the prime in the sum excises the states with $\varepsilon_n = \omega_k$, and $\Gamma_k$ is the decay rate determined by Fermi's Golden Rule which is  strictly valid in the infinite time limit,
\be \Gamma_k = 2\pi\,\rho(k_0=\omega_k,k)= \Bigg\{ \begin{array}{c}
                                                     \frac{Y^2\,M}{8\pi\,\gamma_k} ~~~~\mathrm{Yukawa} \\
                                                     \frac{\lambda^2}{8\pi \,M\,\gamma_k} ~~~~\mathrm{Scalar}\,.
                                                   \end{array}
  \label{Gammafgr} \ee In these expressions $\gamma_k = \omega_k/M$ is the Lorentz factor and in (\ref{Gammafgr}) describes  the usual time dilation factor in the decay rate. The results above are standard, however, there are subtle but important aspects pertaining to the quantum field theory setting. For the super renormalizable case with the spectral density (\ref{rhoS}) the wave function renormalization is \emph{finite} as the spectral density becomes constant at large $k_0$, therefore the integral in (\ref{waf}) is convergent. However, in the renormalizable case with $\rho(k_0,k)$ given by (\ref{rhoR}), the spectral density grows as $k^2_0$ for large $k_0$ and the integral becomes \emph{linearly divergent}. This is a distinct consequence of the renormalizability of the theory with Yukawa coupling:  the spectral density has mass (energy) dimension one, but features a pre-factor  $1/\omega_k$ from the normalization of the decaying state. In the super renormalizable case with $\lambda$ featuring mass dimension one, the prefactor $\lambda^2/\omega_k$ has mass dimension one therefore the remaining contribution to the spectral density is dimensionless and the total integrand for $ {Z}$ falls off as $1/k^2_0$ for large $k_0$. In the renormalizable case $Y$ is dimensionless, therefore the prefactor $Y^2/\omega_k$ requires two extra powers of mass to yield a contribution of mass dimension one, this implies the $k^2_0$ dependence of the spectral density,   ultimately resulting in the divergence of $ {Z}$.

  In the  conventional   renormalization program,   counterterms are introduced in the Hamiltonian,  these  are part of the interaction and are obtained systematically in perturbation theory  by requiring the cancellation of the divergences at a given scale. This familiar renormalization procedure \emph{does not} work in the real time calculation of the survival probability as can be understood from the following argument. Consider adding such a counterterm to the interaction Hamiltonian $\delta H^{ct}_i$ which is second order in the couplings because it must cancel divergences originating from a one loop self energy. Such counterterm, when considered as part of the interaction modifies the first term (linear in $H_I$) in (\ref{UI}). However, the interaction Hamiltonian including the counterterm is hermitian, and yields  a purely imaginary contribution to the diagonal matrix element for the survival \emph{amplitude} (\ref{amp}), which, however,  does \emph{not} contribute to the survival \emph{probability}. Therefore such counterterm   cannot cancel the divergence in $ {Z}$ in the real time evolution of the survival probability in the renormalizable case.

We note that the expression for $ {Z}$ (\ref{waf}) is the same as that obtained in quantum mechanics, and it describes the amplitude of the unperturbed state in the \emph{exact} eigenstate of the Hamiltonian, constructed up to second order in perturbation theory. To highlight this, consider the case when the mass of the Higgs scalar, $M$,  is below the two particle threshold, namely $M< 2m$. In this case the Higgs scalar cannot decay and $\Gamma_k=0$, nevertheless $ {Z} < 1$, and the probability of finding the initial state in the time evolved state at asymptotically long time is $ {Z}$. In the stable case, this is the probability of finding the unperturbed state in the eigenstate of the full Hamiltonian. The corollary of this discussion is that the survival probability (\ref{Poft}) actually describes \emph{two different processes}: the build-up of the \emph{quasiparticle}, encoded in $ {Z}$ and its decay because this quasiparticle state is embedded in the continuum becoming unstable, this is encoded in $\Gamma_k$. In the renormalizable case the second order contribution to $ {Z}$ is ultraviolet divergent. One of our goals is to understand how to meaningfully separate the dynamics of the build-up of the quasiparticle from its decay in a consistent manner.

\subsection{Non-perturbative method}
Consider the  Hamiltonian $H =H_0+H_i$, the time evolution of states $\ket{\Psi(t)} $ in the interaction picture
of $H_0$ is given by
\be i \frac{d}{dt}|\Psi(t)  \rangle_I   = H_I(t)\,|\Psi(t) \rangle_I ,  \label{intpic}\ee
where the interaction Hamiltonian in the interaction picture is
\be H_I(t) = e^{iH_0\,t} H_i e^{-iH_0\,t}\,. \label{HIoft}\ee

The formal solution of the time evolution equation (\ref{intpic}) is given by
\be |\Psi (t)\rangle_I  = U_I(t,t_0) |\Psi(t_0)\rangle_I \label{sol}\ee
where   the time evolution operator in the interaction picture $U_I(t,t_0)$ obeys \be i \frac{d}{dt}U_I(t,t_0)  = H_I(t)U_I(t,t_0)\,. \label{Ut}\ee

Now we can expand \be |\Psi(t)\rangle_I = \sum_n C_n(t) |n\rangle \,,\label{decom}\ee where $H_0|n\rangle=\varepsilon_n\ket{n}$, from eq.(\ref{intpic}) one finds the {\em exact} equations of motion for the amplitudes $C_n(t)$, namely

\be \dot{C}_n(t) = -i \sum_m C_m(t) \langle n|H_I(t)|m\rangle  \,. \label{eofm}\ee The hermiticity of $H_i$ guarantees unitarity, it is straightforward to show that
\be \frac{d}{dt} \sum_n |C_n(t)|^2 = 0 \,.\label{unit}\ee  Consider that initially only one amplitude, $C_1$ is different from zero and all others vanish, the result (\ref{unit}) entails that
\be \sum_n |C_n(t)|^2 = |C_1(0)|^2 \,. \label{sumunit}\ee

\begin{figure}[ht!]
\begin{center}
\includegraphics[height=3in,width=3in,keepaspectratio=true]{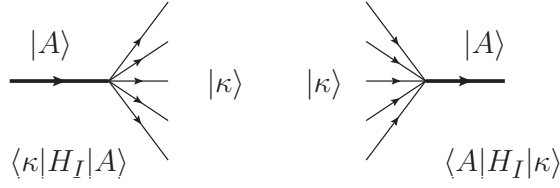}
\caption{Transitions $|A\rangle \leftrightarrow |\kappa\rangle$ in first order in $H_I$.}
\label{fig1:transitions}
\end{center}
\end{figure}

Although   equation (\ref{eofm}) is exact, it generates an infinite hierarchy of simultaneous equations when the Hilbert space of states spanned by $\{|n\rangle\}$ is infinite dimensional. However, this hierarchy can be truncated by considering the transition between states connected by the interaction Hamiltonian at a given order in $H_i$. Thus
consider the situation depicted in figure~\ref{fig1:transitions} where a given state, $|A\rangle$, couples to a set of states $\left\{|\kappa\rangle\right\}$,   and these states couple back only to $|A\rangle$ via $H_i$ to first order in the interaction.

In the cases under consideration   the single particle state of the Higgs field $\ket{1^{\Phi}_{\vk}}$, with amplitude $C_{\Phi}$ and energy $\omega_k=\sqrt{k^2+M^2}$ couples to an intermediate state $|\kappa\rangle$ with a fermion-anti fermion pair in the case of Yukawa coupling, or a pair  bosonic $\chi$ particles in the super renormalizable case. In these cases we have \bea \dot{C}_\Phi(t) & = & -i \sum_{\kappa} \langle 1^{\Phi}_{\vk}|H_I(t)|\kappa\rangle \,C_\kappa(t)\label{CA}\\
\dot{C}_{\kappa}(t) & = & -i \, C_\phi(t) \langle \kappa|H_I(t) |1^{\Phi}_{\vk}\rangle \label{Ckapas}\eea where the sum over $\kappa$ is over all the intermediate states $\ket{\kappa}$ coupled to $|1^{\Phi}_{\vk}\rangle$ via $H_i$.

\begin{figure}[ht!]
\begin{center}
\includegraphics[height=4in,width=4in,keepaspectratio=true]{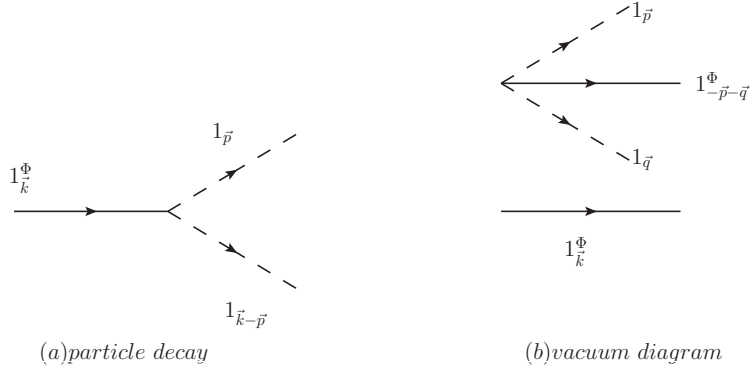}
\caption{Decay and vacuum diagrams for  $ \ket{1_k^{\Phi}}$ to first order in $H_i$. Solid lines single particle states of the the   fields $\Phi $, dashed lines are single particle states of the either the $\chi$ of fermion (anti-fermion) field.}
\label{fig1:coupling}
\end{center}
\end{figure}

Consider the initial value problem in which at time $t=0$ the state of the system $|\Psi(t=0)\rangle = |1^{\Phi}_{\vk}\rangle$, namely \be C_\Phi(0)\neq 0,\   C_{\kappa}(0) =0 .\label{initial}\ee  We   solve eq.(\ref{Ckapas}) and then use the solution in eq.(\ref{CA}) to find \bea  C_{\kappa}(t) & = &  -i \,\int_0^t \langle \kappa |H_I(t')|1^{\Phi}_{\vk}\rangle \,C_\Phi(t')\,dt' \label{Ckapasol}\\ \dot{C}_\Phi(t) & = & - \int^t_0 \Sigma(t,t') \, C_\Phi(t')\,dt' \label{intdiff} \eea where \be \Sigma(t,t') = \sum_\kappa \langle 1^{\Phi}_{\vk}|H_I(t)|\kappa\rangle \langle \kappa|H_I(t')|1^{\Phi}_{\vk}\rangle = \sum_{\kappa} |\bra{1^{\Phi}_{\vk}}H_i\ket{\kappa}|^2\, e^{i(\omega_k-\varepsilon_\kappa)(t-t')}\,.\label{sigma} \ee   Inserting the solution for $C_\Phi(t)$ into eq.(\ref{Ckapasol}) one obtains the time evolution of amplitudes $C_{\kappa}(t)$ from which we can compute  the time dependent probability to populate the state $|\kappa\rangle$ given by $|C_\kappa(t)|^2$. This is the essence of the non-perturbative Weisskopf-Wigner   method\cite{ww,book} which has been extended to quantum field theory in refs.\cite{boyww,boycosww}.  With $|1^{\Phi}_{\vk}\rangle$ being the single particle Higgs state with momentum $\vk$, there are two different types of contributions: a) a disconnected diagram in which the initial single particle state evolves freely with a \emph{three particle} disconnected contribution and b) a connected diagram in which the initial single particle state decays into a pair of particles in the intermediate state. Both contributions are depicted in fig. (\ref{fig1:coupling}).

\begin{figure}[ht!]
\begin{center}
\includegraphics[height=3in,width=4in,keepaspectratio=true]{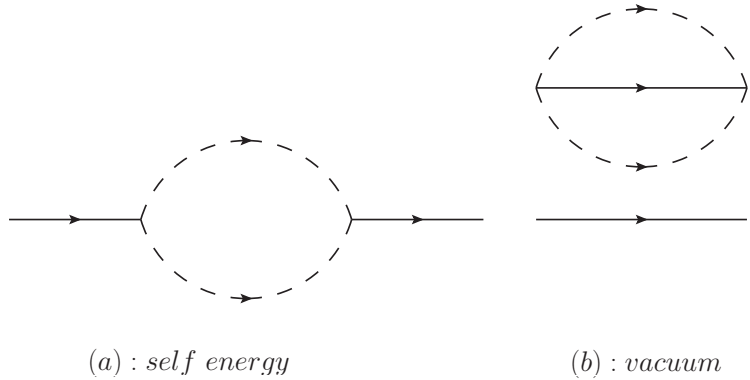}
\caption{Contributions to the self-energy   for decay (a) and vacuum diagram (b) for  $\ket{1^{\Phi}_{\vk}}$ to first order in $H_I$ with the same notation as in fig.(\ref{fig1:coupling}). }
\label{fig3:decayvacuumse}
\end{center}
\end{figure}

These processes yield two different contributions to $\sum_\kappa \bra{1_{\vk}^{\Phi}}H_I(t)\ket{\kappa}
\bra{\kappa}H_I(t')\ket{1_{\vk}^{\Phi}} $, depicted in fig. (\ref{fig3:decayvacuumse}).

The disconnected diagram is a vacuum correction and is absorbed into a redefinition of a single particle state constructed out of the full vacuum state (see ref.\cite{boyww,boycosww,herring}). It is straightforward to see that the amplitude equation  for the vacuum state, is determined by the disconnected vacuum diagram, indeed the amplitude equations for \emph{all} states feature the disconnected contribution. The connected diagram yields the self-energy of the single particle state. Only the connected diagram is relevant for the dynamics of the single particle state and the survival probability of this state. It is straightforward to find  the perturbative result (\ref{Poft}) from equation (\ref{intdiff}) by replacing $C_\Phi(t')=C_{\Phi}(0)=1$ in (\ref{intdiff}) and carrying out the time integrals, and confirm  that the result is given by   eqn. (\ref{Poft}). However, se week a non-perturbative solution of the amplitude equation (\ref{intdiff}).

Introducing the spectral density (\ref{specdens}), the self-energy (\ref{sigma}) can be written as
\be \Sigma(t,t') = \int^{\infty}_{-\infty}\rho(k_0,k) \,e^{i(\omega_k-k_0)(t-t')}\,dk_0 \,,\label{sigrho}\ee where for the renormalizable  (Yukawa) and super renormalizable cases $\rho(k_0,k)$ are given by (\ref{rhoR},\ref{rhoS}) respectively.
Equation (\ref{intdiff}) can be solved by Laplace transform with initial condition $C_{\Phi}(0)$. Definining the Laplace transform
\be \widetilde{C}_{\Phi}(s) = \int^\infty_0 e^{-st}\,C_{\Phi}(t) \,dt \, \label{lapla}\ee  and similarly for the Laplace transform of the self-energy $\widetilde{\Sigma}(s)$, for which  we find upon using (\ref{sigrho})   the result
\be \widetilde{\Sigma}(s) = \int^{\infty}_{-\infty} \frac{\rho(k_0,k)}{s+i(k_0-\omega_k)}\,. \label{tilsig}\ee
Taking the Laplace transform of (\ref{intdiff}) we find
\be \widetilde{C}_{\Phi}(s) = \frac{C_{\Phi}(0)}{s+\widetilde{\Sigma}(s)}\,.\label{laplafiamp}\ee The inverse Laplace transform yields $C_{\Phi}(t)$, namely
  \be C_{\Phi}(t) = \int^{i\infty+\epsilon}_{-i\infty +\epsilon} \frac{ds}{2\pi\,i} ~\widetilde{C}_{\Phi}(s)\,e^{st} \label{invlapla}\ee where the $\epsilon \rightarrow 0^+$ determines the Bromwich contour in the complex $s$-plane parallel to the imaginary axis to the right of all the singularities of $\widetilde{C}_{\Phi}(s)$. Writing $s=i(\omega-i\epsilon)$ we find \be C_{\Phi}(t) = C_{\Phi}(0)\,\int_{-\infty}^{\infty} \frac{d\omega}{2\pi\,i}~ \frac{e^{i\omega t}}{\Bigg[\omega-i\epsilon - \int_{-\infty}^{\infty} dk_0~\frac{\rho(k_0,k)}{\omega+k_0-\omega_k-i\epsilon} \Bigg]  }\,. \label{CAfin}\ee

   In perturbation theory there is a complex pole very near $\omega =0$ which can be obtained directly by
expanding the integral in the denominator near $\omega =0$. We find \be  \int_{-\infty}^{\infty} dk_0~\frac{\rho(k_0,k)}{\omega+k_0-\omega_k-i\epsilon} \simeq -\Delta \omega - z \,\omega + i \,\frac{\Gamma_k}{2} \label{aproxi}\ee where \bea \Delta \omega  & = &   \int_{-\infty}^{\infty} dk_0 \, \mathcal{P} \Bigg(\frac{\rho(k_0,k)}{(\omega_k-k_0)} \Bigg) \label{energyshift} \\ \Gamma_k & = & 2\pi\,\rho(k_0=\omega_k,k) \label{width} \\ z  & = &   \int_{-\infty}^{\infty} dk_0 \, \mathcal{P}\Bigg(\frac{\rho(k_0,k)}{(\omega_k-k_0)^2}\Bigg)\,.\label{smallz}\eea   The term $\Delta \omega$ is recognized as the energy renormalization of the Higgs scalar, while $\Gamma_k $ is seen to be the decay rate as found from Fermi's golden rule and coincides with the perturbative result(\ref{Gammafgr}). The {\em long time} limit of $C_\Phi(t)$ is determined by this complex pole near the origin leading to the asymptotic behavior\footnote{There are long time power law corrections which are not relevant for the discussion here, see refs.\cite{cosa,yukacosa,supercosa,urbano,giunti,ramirez}.} \be C_{\Phi}(t)\simeq {Z}  \, e^{-i\Delta \omega^r \,t}\,e^{- {\Gamma^r_k \,t/2}} \,C_{\Phi}(0) \label{tasi}\ee where
\be  {Z}  = \frac{1}{1+z}\simeq 1-z   \label{wavefunc}\ee is the wave function renormalization constant which is recognized as the same contribution in the perturbative case (\ref{waf}), and
\bea \Delta \omega^r  & = &   {Z} \,\Delta \omega \label{DeltaEr}\\ \Gamma^r_k  & = &  {Z} \,\Gamma_k  \,.\label{GammaAr} \eea

For the renormalizable case (\ref{rhoR})  for $m=0$ and introducing a cutoff $\Lambda$ for the $k_0$ integral  we find
\be z= \frac{Y^2\,M}{16\pi^2\,\omega_k}\,\Bigg[\frac{\Lambda-k}{M}+2\,\frac{\omega_k}{M} \ln\Bigg(\frac{\Lambda-\omega_k}{\omega_k-k} \Bigg) -\frac{M}{\Lambda-\omega_k}-\frac{M}{(\omega_k-k)}\Bigg] \,. \label{lilzm0} \ee

\subsubsection{Markovian approximation.}
In weak coupling there is a wide separation of time scales,   the interaction picture amplitudes $C_{\Phi},C_\kappa$ are \emph{slow} in the sense that they evolve over time scales $\propto 1/Y^2$   or $1/\lambda^2$   because $\Sigma \propto Y^2$  or $\propto \lambda^2$ respectively. These scales are  much longer than those in the self-energy $\simeq 1/\omega_k, 1/\varepsilon_n$.  This allows us to use a Markovian approximation in terms of a
consistent expansion in derivatives of $C_\Phi$. This is implemented by defining \be i\mathcal{E}(t,t') = \int^{t'}_0 \Sigma(t,t'')dt'' \,,\label{Wo}\ee so that \be \Sigma(t,t') = i\frac{d}{dt'}\,\mathcal{E}(t,t'),\quad \mathcal{E}(t,0)=0. \label{rela} \ee Integrating by parts in eq.(\ref{intdiff}) we obtain \be \int_0^t \Sigma(t,t')\,C_\Phi(t')\, dt' = i \mathcal{E}(t,t)\,C_\Phi(t) +i \int_0^t \mathcal{E}(t,t')\, \frac{d}{dt'}C_\Phi(t') \,dt'. \label{marko1}\ee The second term on the right hand side is formally of \emph{fourth order} in $H_i$, since $\mathcal{E}\simeq \Sigma \simeq H^2_i$ and $\dot{C_\Phi} \simeq H^2_i$.

This process can be implemented systematically by subsequent integration by parts resulting in higher order differential equations with higher derivative terms suppressed by   powers of $H^2_i$. Up to leading order in this Markovian approximation the equation eq.(\ref{intdiff}) becomes \be \dot{C}_\Phi(t) = -i \mathcal{E}(t,t) C_\Phi(t)   \label{markovian}\ee with the result \be C_\Phi(t) = e^{-i\int_0^t \mathcal{E}(t',t')dt'}\,C_{\Phi}(0)\,. \label{solumarkov}\ee We emphasize that this result provides a \emph{non-perturbative} resummation of the leading order contributions, namely second order in the perturbative expansion. It is the real time equivalent of the Dyson resummation of self-energy diagrams up to second order in perturbation theory. The result (\ref{solumarkov}) is exact up to second order in the perturbative couplings.  The survival probability is then given by
\be P_k(t)= |C_\Phi(t)|^2 = e^{-\alpha_k(t)}\,|C_\Phi(0)|^2 \,,\label{survimark}\ee where in terms of the representation (\ref{sigrho}) and using (\ref{Wo}) it is straightforward to find that the decay function $\alpha_k(t)$ is given by
\be \alpha_k(t)  = 2\int^{\infty}_{-\infty} dk_0 \, \rho(k_0,k) \,\frac{\Big[1-\cos\big[\big( k_0-\omega_k\big)t\big]\Big]}{(k_0-\omega_k)^2} \,. \label{lilgama} \ee Expanding $e^{-\alpha_k(t)} \simeq 1-\alpha_k(t) +\cdots$ and keeping only the first order term in $\alpha_k(t)$ one recovers the perturbative result (\ref{Poft}).

Using the identity (\ref{iden}) (in the sense of distributions) for the asymptotic long-time limit we find in this limit
\be P_k(t) = e^{-\Gamma_k t}~e^{-2z}\,\, P_k(0)\,, \label{asylimimark}\ee where $\Gamma_k$ and $z$ are given by   (\ref{Gammafgr}) and (\ref{smallz}) respectively. To leading order in perturbation theory it follows that $e^{-2z} \simeq  {Z}^2$ with $ {Z}$ the wave function renormalization (\ref{wavefunc}). Therefore the asymptotic long time limit in the Markov approximation  agrees with the result (\ref{tasi}). There are two main reasons to consider the Markov approximation rather than the Laplace transform solution: i) in the cosmological case, the time evolution of the gravitational background prevents  a solution in terms of Laplace transforms, ii) the Markov approximation allows to study the \emph{full time evolution} whereas the Laplace transform yields the asymptotic long time dynamics relatively straightforwardly in terms of the complex pole closest to the real axis in a Breit-Wigner approximation of the propagator, the early transient  dynamics is much more difficult to extract and in general must be obtained numerically. As we argue below this transient dynamics is very important to understand how to identify the time evolution associated with decay from the build-up of the quasiparticle associated with wave function renormalization. In particular the Markov approximation allows an \emph{analytical} understanding of the early transient dynamics of dressing and renormalization in the case of massless fermions (see below), even in this simpler case extracting this early time transient dynamics from the Laplace transform is very difficult.

With $z$ manifestly positive, the survival probability (\ref{asylimimark}) features \emph{two} different damping contributions, the decay secular in time with rate $\Gamma_k$ but also that of the wave function renormalization, which in the case of a renormalizable theory diverges with the cutoff because $\rho_R(k_0,k) \simeq k^2_0$ for large $k_0$. Our goal is to study the total time evolution of the survival probability and understand how to systematically treat the effect of wave function renormalization.

The   time dependence   of $\alpha_k(t)$ can be made more explicity by the following steps: i) we introduce an upper cutoff $\Lambda$ in the $k_0$ integration variable, ii) we change integration variable from $k_0$ to a dimensionless variable $x$ defined  by $k_0 = \omega_k + x/t$.

For the super renormalizable case with spectral density (\ref{rhoS}) we find
\be \alpha_k(t) = \frac{\lambda^2}{8\pi^2\,M^2\,\gamma_k}\,J(t,k) \,, \label{superenalfa} \ee where
\be J(t,k) = Mt\, \int^{X(t)}_{X_{th}(t)}\,\Bigg[1-\frac{4m^2}{M^2\,f(x,t)}\Bigg]^{1/2} \,\frac{\Big[ 1- \cos(x)\Big]}{x^2} ~dx ~~;~~ (\mathrm{super-renormalizable})\,, \label{Jota}\ee with
\be X(t) = (\Lambda-\omega_k)t~~;~~ X_{th}(t) =  -\Big[\gamma_k - \Big(\gamma^2_k+\frac{4m^2}{M^2}-1\Big)^{1/2} \Big]Mt ~~;~~  f(x,t)= 1+ 2\,\frac{\gamma_k\,x}{M t}+\frac{x^2}{M^2 t^2} \,, \label{Xfdefs}\ee where, again $\gamma_k= \omega_k/M$ is the Lorentz factor.

We now focus on the renormalizable case with spectral density (\ref{rhoR}) for which following the same steps and notation as above, we find

\be \alpha_k(t) = \frac{Y^2}{8\pi^2\, \gamma_k}\,\Big[I_1(t,k)+I_2(t,k)+I_3(t,k)\Big]~~(\mathrm{renormalizable})\,, \label{lilgam2}\ee where
\be I_1(t,k) = M\,t\, \int^{X(t)}_{X_{th}(t)}\,\Bigg[1-\frac{4m^2}{M^2\,f(x,t)}\Bigg]^{3/2} \,\frac{\Big[ 1- \cos(x)\Big]}{x^2} ~dx\,, \label{I1f}\ee
\be I_2(t,k) = 2\,\gamma_k\, \int^{X (t)}_{X_{th}(t)}\,\Bigg[1-\frac{4m^2}{M^2\,f(x,t)}\Bigg]^{3/2} \,\frac{\Big[ 1- \cos(x)\Big]}{x} ~dx\,, \label{I2f}\ee
\be I_3(t,k) = \frac{1}{M t}  \int^{X(t)}_{X_{th}(t)}\,\Bigg[1-\frac{4m^2}{M^2\,f(x,t)}\Bigg]^{3/2} \, {\Big[ 1- \cos(x)\Big]}  ~dx\,, \label{I3f}\ee

For $\Lambda \rightarrow \infty$ it follows that $I_1(t,k)$ is finite, $I_2(t,k)$ is logarithmically divergent and $I_3(t,k)$ is linearly divergent in $\Lambda$.

The $x$ integrals  must be carried out numerically in the general case. However, we can provide an analytic form for $m=0$ which highlights the main features of the transient dynamics as well as the main aspects associated with the divergent wave function renormalization. Setting $m=0$ in $\rho_R(k_0,k)$ given by (\ref{rhoR})   we find for the Yukawa (renormalizable) case

\be I_1(t,k) = M\,t \Bigg[Si[(\Lambda-\omega_k)\,t]+Si[(\omega_k-k)\,t]- \frac{\Big[ 1- \cos[(\Lambda-\omega_k)t]\Big]}{(\Lambda-\omega_k)\,t}  \Bigg] - \frac{\Big[ 1- \cos[(\omega_k-k)t]\Big]}{(\omega_k-k)\,t}  \Bigg]\label{I1}\ee

\be I_2(t,k) = 2\,\gamma_k\, \Bigg[ \ln\Big[\frac{\Lambda-\omega_k}{\omega_k-k} \Big]-Ci[(\Lambda-\omega_k)\,t ] + Ci[(\omega_k-k)\,t ]
\Bigg] \label{I2}\ee

\be I_3(t,k) = \frac{(\Lambda-k)}{M}   - \frac{\sin[((\Lambda-\omega_k)\,t]}{M t} - \frac{\sin[((\omega_k-k)\,t]}{M t} \,,\label{I3}\ee where $Si,Ci$ are the sine and cosine integrals respectively with asymptotic limits $Si[x]\rightarrow \pi/2~;~Ci[x] \rightarrow 0$ for $x\rightarrow \infty$. In the limit $\Lambda \gg \omega_k$ and $ (\omega_k -k)t \gg 1$ one finds
\be I_1(t,k) \rightarrow \pi\,M t- \frac{M}{\Lambda-\omega_k}-\frac{M}{\omega_k-k} ~~;~~I_2(t,k) \rightarrow  2\,\gamma_k\,   \ln\Big[\frac{\Lambda-\omega_k}{\omega_k-k} \Big]~~;~~ I_3(t,k) \rightarrow \frac{(\Lambda-k)}{M}\,.\label{lonti}\ee

\begin{figure}[ht!]
\begin{center}
\includegraphics[height=4in,width=4.2in,keepaspectratio=true]{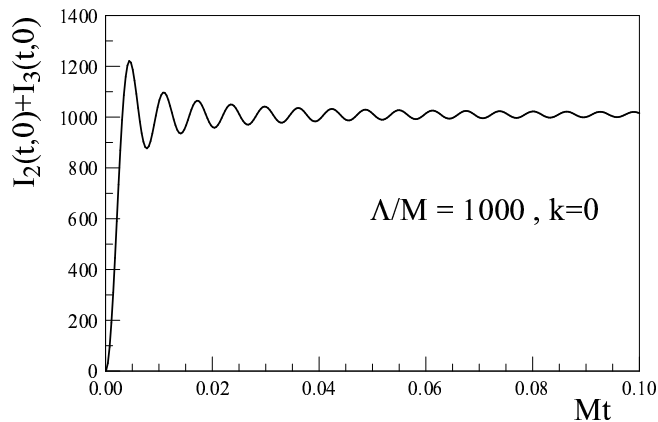}
\caption{$I_2(t,0)+I_3(t,0)$   vs. $Mt$ for $k=0, \Lambda/M=1000$, $m=0$. The results are indistinguishable for any $m\neq 0$.  }
\label{fig:i2plusi3}
\end{center}
\end{figure}

 For the super renormalizable case with $\alpha_k(t)$ given by (\ref{superenalfa},\ref{Jota}) with $m=0$, we find     $J(t,k)=I_1(t,k)$. An important feature of these results is that only $I_1$ grows secularly with time and this contribution is \emph{ultraviolet finite}. In contrast, both $I_2,I_3$ asymptote to  an ultraviolet divergent constant at long time.  Furthermore, $I_2,I_3$ raise to become of order $\simeq \ln[\Lambda]$ and $ \simeq \Lambda$ respectively on very short time scales $t \lesssim 1/\Lambda$, whereas $I_1$ evolves on much longer time scales $t \geq 1/M$.  Figure (\ref{fig:i2plusi3}) shows $I_2(t,0)+I_3(t,0)$ for $m=0$ although the results for $m\neq 0$ are indistinguishable. This figure highlights that these contributions rise nearly to their asymptotic value $\simeq \Lambda$ on a time scale $t \lesssim 1/\Lambda$, also revealing the  oscillatory behavior with a long time tail $\propto 1/t$.

  The time independent terms in the long time limit add up to   $z$  given by (\ref{lilzm0}) and determine the wave function renormalization. The contributions from $I_2,I_3$ which are manifestly divergent as $\Lambda \rightarrow \infty$ are only present for the renormalizable case. This ultraviolet divergence is obviously independent of $m$. The expressions above make explicit the fact that the ultraviolet divergent contributions are \emph{independent of time}, and in the long time limit one finds $\alpha_k(t) \rightarrow \Gamma_k\,t + 2 z+\mathcal{O}(1/t)$, where $z$ is given by the time independent and ultraviolet divergent factors.

  This analysis makes clear that the contributions $I_{2,3}$ which evolve on short time scales $t \lesssim 1/\Lambda$ are associated with the build-up of the quasiparticle, namely the ``bare'' particle dressed by many body correlations,  whereas $I_1$ describes its decay  which for weak coupling occurs  on much longer time scales $t \gg  1/M $.

For $\Lambda \gg \omega_k$ it is straightforward to find that for $t \lesssim 1/\Lambda$ in the renormalizable case
\be \alpha_k(t) \simeq \frac{t^2}{t^2_Z} ~~;~~ t^2_Z = \frac{96\,\pi^2 \omega_k}{Y^2\,\Lambda^3} \,, \label{zeno} \ee and during the time scale for build-up of the quasiparticle the decay law of the survival probability   reflects the \emph{Zeno} effect\cite{naka,review,zenopasca,faccipasca,facchi}
\be |C_{\Phi}(t)|^2 \simeq e^{-t^2/t^2_Z}\,|C_{\Phi}(0)|^2 \, , \label{zenolaw}\ee with the Zeno time scale $t_Z$. A projective measurement of the Higgs state for time scales $t \lesssim 1/\Lambda$ would reveal this suppressed decay law at early times and would witness the formation of the quasiparticle. Although it is very unlikely that a particle physics experiment would reveal a Zeno behavior, such early time transient has been observed in transitions of $Be$ atoms\cite{wineland} and in cold sodium atoms\cite{zenoobs}.

At the end of the Zeno stage for $t\simeq 1/\Lambda$ it follows that $\alpha_k(t) \simeq Y^2 \Lambda/96\pi^2 \omega_k$, reflecting the ultraviolet divergence associated with the wave function renormalization.  In contrast,  in the superrenormalizable case a similar Zeno decay law emerges for $t\lesssim 1/\Lambda$ but with $t^2_Z \simeq \omega_k/\lambda^2 \Lambda$, therefore at the end of the Zeno regime in this case one finds $\alpha \simeq \lambda^2/\omega_k \Lambda$, hence negligible. This is an important distinction between the super renormalizable and renormalizable cases.

The quantum zeno time of the $2-P \rightarrow 1-S$ transition of the hydrogen atom has been obtained in ref.\cite{facchizeno} and found to be $t_Z \approx 1/\Lambda$, the discrepancy with the result (\ref{zeno}) in the renormalizable case can be traced to   very different transition matrix elements and spectral density, where for the relativistic Yukawa quantum field theory the spectral density grows as $\Lambda^3$ whereas in the study of ref.\cite{facchizeno} it grows as $\Lambda^2$.

The large suppression of  the survival probability in the renormalizable case,  a consequence of the formation of the quasiparticle,  requires a physical  renormalization procedure that includes the dynamics of quasiparticle formation. This quasiparticle state ``dressed'' with the many body correlations, should then be taken as the physical unstable state which then decays on longer time scales. One of our main goals is to systematically separate these different physical processes. A simple time \emph{independent} subtraction of $\alpha_k(t)$ of  the divergent contributions is not the correct procedure  because the time evolution from the state initialized at $t=0$ implies  that $\alpha_k(t\rightarrow 0) =0$.

\section{The birth of the quasiparticle}

The ultraviolet divergence of the wave function renormalization in the renormalizable case arises from the divergent contributions to  $I_2(t,k),I_3(t,k)$. These terms describe the \emph{formation of the quasiparticle state} from the bare state by the many body correlations including high energy states up to the cutoff. Figure (\ref{fig:i2plusi3}) shows that $I_2+I_3$ raises up to the cutoff value on time scales $t \simeq 1/\Lambda$ and approach a constant $\propto \Lambda$  for $t \gg 1/\Lambda$ as shown in figure (\ref{fig:i2plusi3}).

There emerge two different  and widely separated time scales: a short time scale $\approx 1/\Lambda$ over which the many body correlations ``dress'' the bare single particle state into a renormalized \emph{quasiparticle}, and a long time scale over which the quasiparticle \emph{decays} $\tau \geq  1/Y^2 M$. Accordingly, the function $\alpha_k(t)$ in (\ref{survimark}) describes these two processes together, suggesting a split of the form
\be \alpha_k(t) = \alpha_{k,d}(t)+\alpha_{k,r}(t)\,. \label{gamasplit}\ee This split is justified by the wide separation in time scales:  $\alpha_{k,d}(t)$ is ultraviolet finite and describes the decay, it evolves slowly in time and  its main contribution is from $I_1(t,k)$. In the renormalizable case the contribution  $\alpha_{k,r}(t)$ is ultraviolet divergent,    it features a very fast time dependence and  approaches a constant $\propto \Lambda$ for $t \gg 1/\Lambda$, its main contribution is from $I_2(t,k)+I_3(t,k)$.  This split can be implemented by  writing  the probability (\ref{survimark}) as
\be |C_{\Phi}(t)|^2 = e^{-\alpha_{k,d}(t)}\,{Z}^{\,2}(t)\,|C_{\Phi}(0)|^2 ~~;~~  {Z}^{\,2}(t) = e^{-\alpha_{k,r}(t)} \,,\label{probsplit}\ee where $ {Z}(t)$ is a \emph{dynamical} wave function renormalization that describes the build-up of the quasi particle and approaches a constant for $t \gg 1/\Lambda$. However, because $\alpha_k(0)=0$ and  the presence of the long time tail $\propto 1/t$ in   $I_{1,2,3}$, there is no  unique manner to separate the divergent from the finite contributions at a fixed, finite time because one can include ultraviolet finite contributions into ${Z}(t)$.   Furthermore, in order to exhibit a decay law directly without mixing the dynamics with the build-up of the quasiparticle we insist that $ {Z}$ be time independent.  The wide separation of time scales between  the build-up of the quasiparticle and its decay allows to introduce an intermediate scale, $t_b$,   the time of ``birth'' of the quasiparticle, so that $1/\Gamma_k \gg t_b \gg 1/\Lambda$ and  for $t> t_b$ we write
\be \alpha_k(t) =  \alpha_k^{(m)}(t;t_b) + \alpha_k(t_b) ~~;~~\alpha_k^{(m)} (t;t_b) = (\alpha_k(t)-\alpha_k(t_b)) \label{minimal} \ee and
\be |C_{\Phi}(t)|^2 = e^{-\alpha_k^{(m)}(t;t_b)}\, \,|C_{\Phi}(t_b)|^2 \,,\label{miniprob}\ee where
\be |C_{\Phi}(t_b)|^2 \equiv  Z_m^{\,2} \,|C_{\Phi}(0)|^2 ~~;~~ Z_m^{\,2}  = e^{-{\alpha}_k(t_b)} \,.
\label{Zetam}\ee Choosing $t_b \gg 1/\Lambda$ the subtraction in (\ref{minimal}) cancels the ultraviolet divergent terms in $\alpha_k(t)$ and $ {Z}_m$ becomes the ultraviolet divergent wave function renormalization. We  identify $|C_{\Phi}(t_b)|^2$ as the probability of the \emph{quasiparticle}, which now describes the \emph{initial state} at time $t_b$. We refer to this manner of isolating the divergent terms as the \emph{minimal subtraction scheme}.  The physical picture of this procedure is clear: taking the scale $1/\Lambda\ll t_b \ll 1/\Gamma_k$ allows the transient dynamics to dress the initial bare state into the quasiparticle state   by many body correlations but on time scales during which this quasiparticle state did not have time to decay, namely $\Gamma_k t_b \ll 1$. In weakly coupled theories with a large cut-off there is a wide range of scales $t_b$ that can be suitably chosen within this window.

Obviously $|C_{\Phi}(t)|^2$ does not depend on the (arbitrary) scale $t_b$ so that the probability obeys a \emph{renormalization group equation}
\be \frac{d}{dt_b} |C_{\Phi}(t)|^2 =0 \,. \label{rgeqn}\ee

This identity is simply the statement that the total dynamical evolution is independent of the particular time scale $t_b$ at which the quasiparticle is defined. Thus this time scale at which the divergences in the quasiparticle formation are absorbed plays the same role as the energy-momentum scale at which renormalized  couplings are defined in a renormalizable theory.

\begin{figure}[ht!]
\begin{center}
\includegraphics[height=5in,width=5in,keepaspectratio=true]{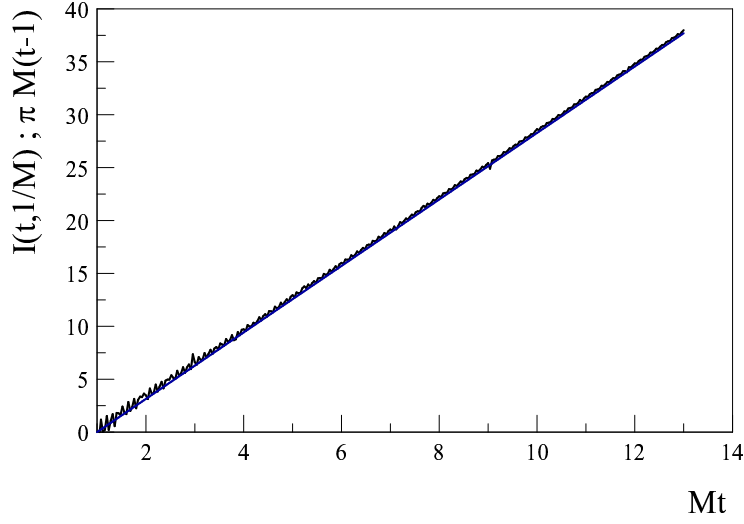}
\caption{$I_0(t,t_b)~;~ \pi(Mt-1)$   vs. $Mt$ for $k=0, \Lambda/M=1000, t_b =1/M, m=0$.   }
\label{fig:it1}
\end{center}
\end{figure}

Following the definition (\ref{lilgam2}) in this minimal subtraction scheme the effective decay function is

\be \alpha^{(m)}_k(t;t_b) = \frac{Y^2\,M}{8\pi^2\,\omega_k}\,I_k(t;t_b)~~;~~ I_k(t;t_b) = \sum_{j=1}^{3} \Big(I_j(t,k)-I_j(t_b,k) \Big)\,. \label{tildegam2}\ee Fig. (\ref{fig:it1}) shows $I_0(t,1/M)$ and  $\pi(Mt-1)$ vs. $Mt$ for $k=0,\Lambda/M =1000,t_b=1/M, m=0$. The rapid early time oscillations feature frequencies of the order of the cutoff and  reflect the remnants of the build-up of the quasiparticle, as $t_b$ is diminished, the amplitude of these oscillations increases in accord with the $\propto 1/t$ long time fall off of these terms explicit in $I_3$ (\ref{I3}). Conversely, larger $t_b$ results in smaller amplitude of the early time oscillatory terms, reflecting, again the $1/t$ tails in these contributions.

We can introduce another subtraction scheme that removes the rapid oscillations which are the remnant of the quasiparticle formation.  For this we introduce
\be I_{D}(t,k;t_b) \equiv  (I_1(t,k)-I_1(t_b,k)) ~~;~~ I_{Q}(t,k;t_b)= I_2(t,k)+I_3(t,k)+I_1(t_b;k) \label{Isplit} \ee and
\be \alpha_{k,d}(t;t_b) = \frac{Y^2\,M}{8\pi^2\,\gamma_k}\,I_D(t,k;t_b) ~~;~~ \alpha_{k,q}(t;t_b) = \frac{Y^2\,M}{8\pi^2\,\gamma_k}\,I_Q(t,k;t_b) \,, \label{gammaSQ} \ee
so that
\be |C_{\Phi}(t)|^2 = e^{-\alpha_{k,d}(t;t_b)}\,Z^{\,2}_Q(t;t_b)\,|C_{\Phi}(0)|^2 ~~;~~ Z^{\,2}_Q(t;t_b) = e^{-\alpha_{k,q}(t;t_b)}\,. \label{onshell}\ee Again the above result is independent of the time scale $t_b$. The time dependent wave function renormalization $Z_Q(t,t_b)$ features the ultraviolet divergences and saturates to a constant value for $t \gg 1/\Lambda$.  Figure (\ref{fig:i2plusi3}) clearly shows that for $t_b\gg 1/\Lambda$ the contribution from $I_2+I_3$ oscillates with diminishing amplitude ($\propto 1/t$) around its asymptotic value $\simeq \Lambda$.  Taking $t_b$ to be an intermediate scale with $1/\Gamma_k \gg t_b \gg 1/\Lambda$  that determines the quasiparticle formation,   it follows that   $Z_Q(t,t_b) \simeq Z_Q(t_b,t_b)$     for $t > t_b$,  so that
\be    |C_{\Phi}(t)|^2 = e^{-\alpha_{k,d}(t;t_b)}\,|C_{\Phi}(t_b)|^2 ~~;~~ |C_{\Phi}(t_b)|^2 = Z^{\,2}_Q(t_b;t_b)\,|C_{\Phi}(0)|^2 ~~;~~ t> t_b \,, \label{Qevol}\ee identifying $C_{\Phi}(t_b)$ as the quasiparticle amplitude. Equation (\ref{Qevol}) describes the decay of the quasiparticle state on time scales much longer than its formation. We refer to this alternative manner of absorbing the ultraviolet divergences as the \emph{on-shell scheme} because $\alpha_{k,d}(t;t_b)$ coincides at long time with the result from Fermi's Golden Rule,   it only includes $I_1$, namely the contribution to the spectral density evaluated on the mass shell.

 \begin{figure}[ht!]
\begin{center}
\includegraphics[height=4in,width=4.2in,keepaspectratio=true]{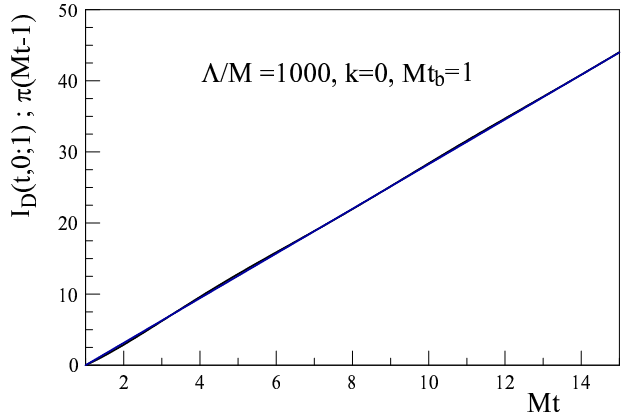}
\caption{$I_S(t,k,t_b)~;~ \pi(Mt-1)$   vs. $Mt$ for $k=0, \Lambda/M=1000, t_b =1/M, m=0$. The two curves are indistinguishable.  }
\label{fig:onshell}
\end{center}
\end{figure}

 Figure (\ref{fig:onshell}) shows $I_D(t,k;t_b)$ and  $\pi(Mt-1)$ for $\Lambda/M =1000,k=0, Mt_b=1$ vs. $ Mt$, the curves are indistinguishable, and the oscillations associated with the remnant of the quasiparticle have disappeared, now being completely absorbed into the wavefunction renormalization. Figure (\ref{fig:i2plusi3}) clearly shows that for $Mt_b \simeq 1$ the contribution from $I_2+I_3$ has reached its asymptotic limit. Obviously the on-shell and minimum subtraction schemes only differ by finite (albeit strongly oscillatory) terms.

While it is very unlikely that particle physics experiments would reveal the transient dynamics of the formation of the renormalized state, the ``birth'' of a quasiparticle state has been observed in ultrafast experiments in silicon\cite{birth} where the phonon states are dressed by electron-hole pairs, a situation very similar to the scalar Higgs field ``dressed'' by fermion-antifermion pairs.

\section{Delayed decay for ultrarelativistic particles.}

The effect of a Lorentz boost enters in \emph{two ways} in the decay functions (\ref{superenalfa}), \ref{lilgam2}). First in the denominator of the prefactors proportional to the coupling, this is the usual time dilation effect, but also in $X_{th}(t)$,  the lower limit of the integrals in (\ref{superenalfa}) and (\ref{I1f})   given by eqn.  (\ref{Xfdefs}). This is a new effect, which becomes explicit only by studying the decay in real time including transient dynamics. After absorbing the contributions from $I_{2,3}$ into the wave function renormalization as discussed above, we consider    the contribution from $I_1$ only. Taking $\Lambda t \gg 1$ and focusing on the lower limit in these integrals, in the limit $t\rightarrow \infty$ (and for $M> 2m$), $X_{th}(t) \rightarrow -\infty$ and the long time limit is determined by the linear secular growth in time, with the coefficient given by Fermi's Golden rule or, equivalently, the S-matrix calculation of the decay rate. It is clear from the expression for $X_{th}(t)$ (in eqn. (\ref{Xfdefs})) that for $\gamma_k \gg 1$ the lower limit of    $I_1(k,t)$ which features secular growth (similarly for $J(k,t)$) lingers near $X_{th}(t) \simeq 0$ during time scales $Mt \lesssim \Big[\gamma_k - \Big(\gamma^2_k+\frac{4m^2}{M^2}-1\Big)^{1/2} \Big]^{-1} $. For $\gamma_k \gg 1$ this becomes a very long time scale, during which the lower limit remains near $X_{th}(t)\simeq 0$ and the value of the integral in (\ref{I1f}) is diminished by almost a factor $1/2$ with respect to the rate in the long time limit, because only the positive values of $x$ contribute. An explicit example of this behavior is gleaned from the first two terms in $I_1(t,k)$ for the $m=0$ case given by  (\ref{I1}). These two terms determine the linear secular growth of $\alpha_k(t)$. The function $Si[x]$ features the behavior $Si[x\rightarrow 0] \rightarrow 0$, and $Si[x\rightarrow \infty] \rightarrow \pi/2$ rising to its asymptotic value for $x \simeq \pi$. For $\Lambda t\rightarrow \infty$ the first $Si[(\Lambda-\omega_k)\,t]$
 function in (\ref{I1}) yields a factor $\pi/2$ but the argument of the second $Si[(\omega_k-k)\,t]$ function behaves as $Mt/\gamma_k$ as the Lorentz factor $\gamma_k \gg 1$. Therefore this second term yields a vanishing contribution up to a time scale $Mt \approx \pi\,\gamma_k$, thereby \emph{suppressing} the coefficient of the linear secular term by a factor $\approx 1/2$ during this time interval. Whereas in the very long time limit $t \gg 1/(\omega_k-k)$ the sum $Si[(\Lambda-\omega_k)\,t]+Si[(\omega_k-k)\,t] \rightarrow \pi$ the second $Si$ function is vanishingly small during a time interval $t \lesssim 1/(\omega_k-k)$, thereby suppressing the rate by a factor $\approx 1/2$.

\begin{figure}[ht!]
\begin{center}
\includegraphics[height=4in,width=5.2in,keepaspectratio=true]{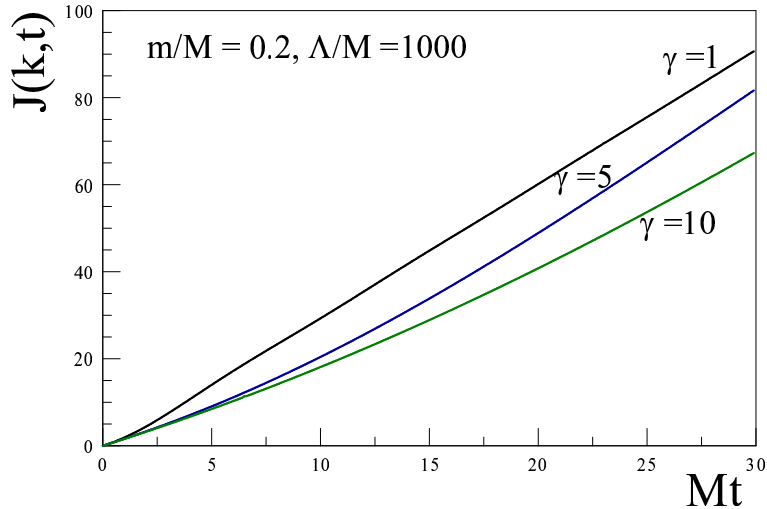}
\caption{Super renormalizable case: $J(t,k)~$   vs. $Mt$ for $\gamma_k=1,5,10; \Lambda/M=1000$.    }
\label{fig:boostsup}
\end{center}
\end{figure}

\begin{figure}[ht!]
\begin{center}
\includegraphics[height=4in,width=5.2in,keepaspectratio=true]{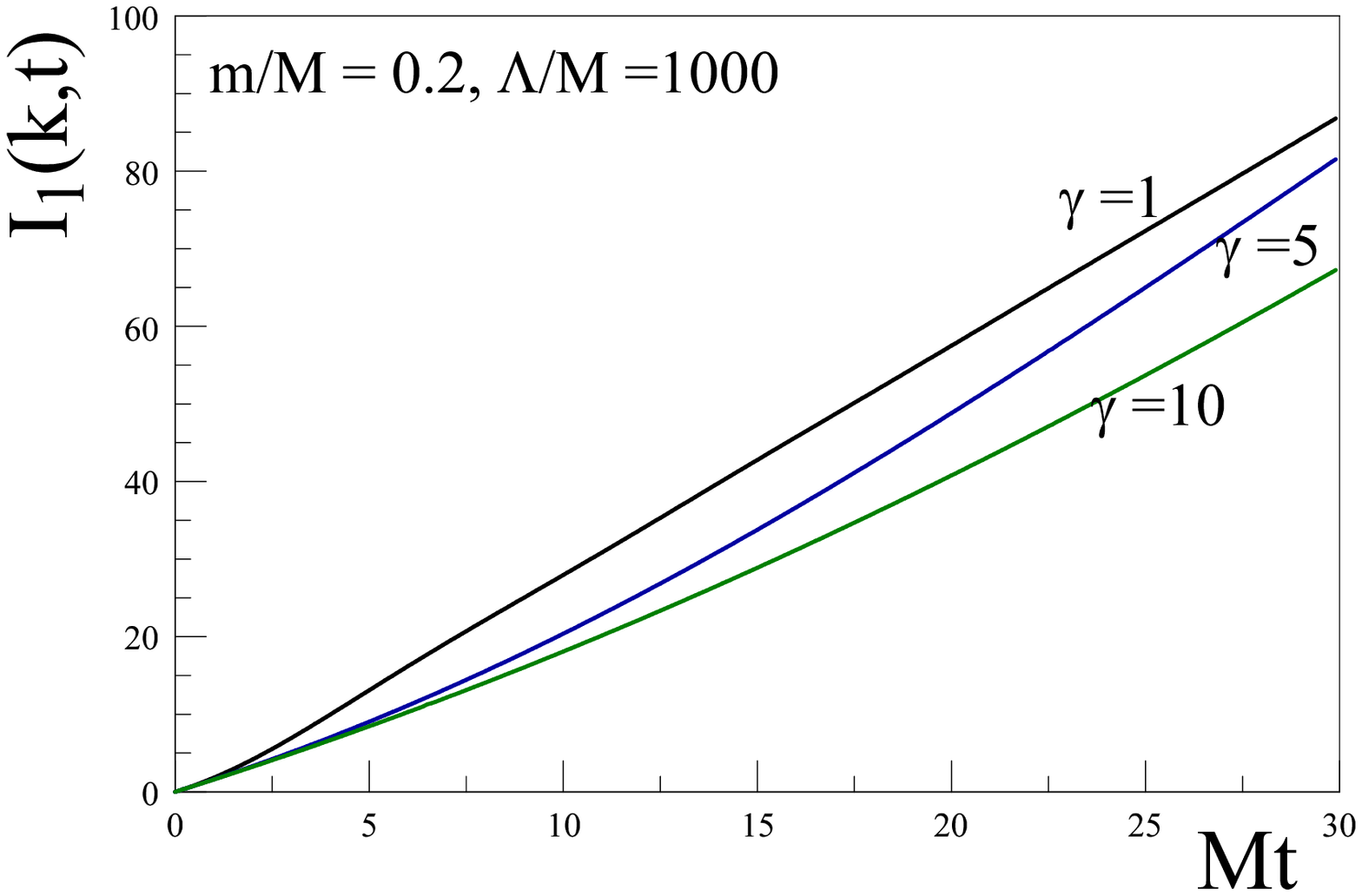}
\caption{Renormalizable case: $I_1(t,k)~$   vs. $Mt$ for $\gamma_k=1,5,10; \Lambda/M=1000$.    }
\label{fig:boostren}
\end{center}
\end{figure}

 This phenomenon  results in a \emph{delayed decay} quite different from the usual time dilation   manifest in the prefactors in (\ref{superenalfa}) and  (\ref{lilgam2}). This behavior is displayed in figures (\ref{fig:boostsup}) and (\ref{fig:boostren}) for the super renormalizable and renormalizable (Yukawa) cases respectively.

 These figures confirm the suppression by a factor $\approx 1/2$ for large Lorentz boost factor $\gamma_k$ with respect to the case of $\gamma_k \simeq 1$, as well as the delayed rise of the secularly growing contribution to $I_1$.  The physical reason behind this suppression can be traced back to the narrowing of the phase space for large boosts: the energy of the decaying particle $\omega_k$ becomes very close to the threshold (from above) at $\sqrt{k^2+4m^2}= \sqrt{\omega^2_k+4m^2-M^2}$ as $\gamma_k \rightarrow \infty$, thereby diminishing the available phase space for decay. This phenomenon is clearly different from the increase in the lifetime by time dilation.

 In the asymptotic long time limit all the curves merge with the $\gamma_k=1$ case indicating that the delay from large Lorentz boost of  the functions $J,I_1$ is a transient phenomenon, which, however, lasts a long interval of time for $\gamma_k \gg 1$. This extra increase in the lifetime, is obviously very different from that arising from time dilation and cannot be described within the S-matrix formulation  because it is exclusively a transient phenonemon, which is not revealed when taking the $t\rightarrow \infty$ limit.

\section{Antizeno effect: accelerated decay by uncertainty}\label{sec:antizeno}

The decay function $\alpha_k(t)$ given by (\ref{lilgama}) involves the integral of the spectral density $\rho(k_0,k)$ with the function
\be C(k_0,t) =  \frac{\Big[1-\cos\big[\big( k_0-\omega_k\big)t\big]\Big]}{(k_0-\omega_k)^2} \,. \label{C0}\ee This function features a maximum at $k_0 = \omega_k$ with half-width $\approx 2.75/t$ which determines the energy uncertainty at a given time. This uncertainty brings the interesting possibility of probing higher energy thresholds during a finite interval of time, with a concomitant enhancement of the decay function. To illustrate this phenomenon, let us consider the case of two fermionic species with   masses $m_2 > m_1$ coupled to the Higgs scalar field $\Phi$ and assume that $2m_2> M > 2m_1$. In this case the scalar particle is above the threshold to decay into two fermions of mass $m_1$, but below the threshold to decay into the heavier species with mass $m_2$.  The total spectral density is given by
\bea \rho_R(k_0,k) = &&\frac{Y^2_1}{16\pi^2 \,\omega_k}\, (k^2_0-k^2) \Bigg\{ \Bigg[1-\frac{4m^2_1}{k^2_0-k^2}\Bigg]^{3/2}\,~~\Theta(k_0)\,\Theta(k^2_0-k^2-4m^2_1) \nonumber \\
&+& y^2\, \Bigg[1-\frac{4m^2_2}{k^2_0-k^2}\Bigg]^{3/2}\,~~\Theta(k_0)\,\Theta(k^2_0-k^2-4m^2_2)
 \Bigg\} \,, \label{rhoRtotal}\eea where
 \be y = \frac{Y_2}{Y_1} \,. \label{ydef}\ee If $\Phi$ is the Higgs field the ratio of Yukawa couplings is given by the mass ratio, namely $y = m_2/m_1 >1$.

 If the condition
 \be \omega_k + \frac{2.75}{t} > \sqrt{\omega^2_k + 4m^2_2- M^2} \,,\label{condiaz} \ee is fulfilled then the function $C(k_0,t)$ has a non-vanishing overlap with the spectral density corresponding to the higher threshold. In this case the $k_0$ integral receives a contribution from the second term in (\ref{rhoRtotal}), thereby enhancing the decay function and accelerating the decay of the (quasi) particle. This is the essence of the anti Zeno effect\cite{antizenokuri,review,lewantizeno}: the energy uncertainty at the finite time $t$ allows to sample a larger portion of the spectral density, in this case that associated with the higher mass threshold, leading to an  enhancement of the decay function.

 \begin{figure}[ht!]
\begin{center}
\includegraphics[height=4in,width=4.5in,keepaspectratio=true]{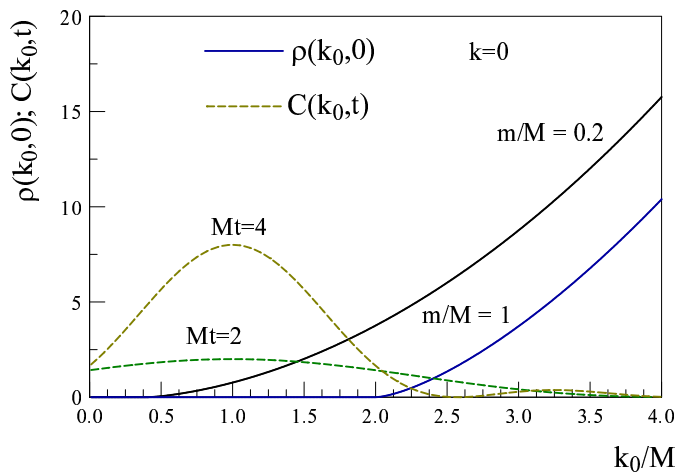}
\caption{Antizeno effect for the renormalizable case for $k=0$. Dashed lines correspond to the function $C(k_0,t)$ (\ref{C0}) and solid lines to the spectral density for the cases of below  ($2m/M=0.2$0) and above  ($2m/M = 2$) threshold respectively  vs. $k_0/M$ for $Mt=2,4$ and $k=0$.    }
\label{fig:antizenoren}
\end{center}
\end{figure}

This phenomenon is displayed in figure (\ref{fig:antizenoren}) which shows the function $C(k_0,t)$ and the spectral densities corresponding to a low and a high mass threshold for $k=0$. Note the overlap of $C(k_0,t)$ with the spectral density with the higher mass threshold. The overlap diminishes at larger time as a consequence of diminishing the energy uncertainty. At early times the function $C(k_0,t)$ probes a larger region of the higher mass threshold (the value $Mt=2$ in fig. (\ref{fig:antizenoren})). As time increases the function $C(k_0,t)$ becomes narrower and well localized below the high mass threshold, effectively closing this transient decay channel.

For the case when $\Phi$ is the Higgs scalar, the Yukawa couplings are proportional to the mass of the intermediate fermion.  Therefore, in this case the larger Yukawa coupling yields a further enhancement of the antizeno effect  as higher mass threshold receives  a proportionally   larger contribution to the total spectral density.

The condition for a substantial enhancement of the decay function from the anti Zeno effect (\ref{condiaz}) can also be written as
\be  2.75\,\Bigg[\Big(\gamma^2_k + \frac{4m^2_2}{M^2}- 1\Big)^{1/2}-\gamma_k   \Bigg]^{-1}>{Mt}  \,,\label{condiaz2} \ee  this is the statement that the half-width of the function $C(k_0,t)$ is larger than the distance between the ``pole'' and the higher mass threshold.  For $\gamma_k \gg 1$ it follows that this condition becomes
\be \frac{2.75 \,\gamma_k}{\Big(\frac{4m^2_2}{M^2}-1\Big)} > Mt\,,  \label{aZlargegam}\ee clearly indicating that the condition will be fulfilled for larger intervals of time for large Lorentz boost factor.  This phenomenon, again has a simpler interpretation, the relative distance between the pole at $\omega_k$ and the higher mass threshold at $\sqrt{\omega^2_k+ 4m^2_2-M^2}$ becomes \emph{smaller} for larger Lorentz factor, therefore the energy uncertainty at a given time $t$ probes larger regions of the higher mass contribution of the spectral density, thereby enhancing the anti-Zeno effect and the increase in the decay function.

The decay channel into the higher mass states remains open by the energy uncertainty during the time interval during which the condition (\ref{condiaz2}) (or (\ref{aZlargegam})) is fulfilled. As time evolves this channel effectively closes at a time scale $t_c$ with

\be  Mt_c \simeq   2.75\,\Bigg[\Big(\gamma^2_k + \frac{4m^2_2}{M^2}- 1\Big)^{1/2}-\gamma_k   \Bigg]^{-1}   \, \label{closing} \ee when decay into the heavier mass states stops. For time scales $t\gg t_c$ the decay function approaches the linear time behavior with the rate given by the S-matrix calculation, however this asymptotic behavior emerges at very long times when the (quasi) particle has already decayed perhaps substantially depending on the values of couplings and Lorentz factor.

We now study this phenomenon numerically to confirm this analysis.  Performing the same change of variables leading to (\ref{lilgam2}-\ref{I3f}) and keeping solely the contributions to $I_1(t,k)$ after renormalization, we now define
 \be I^{tot}_1(t,k) = I_1(t,k;m_1)+ y^2\,I_1(t,k;m_2) \label{Itotal}\ee where the contributions $I_1(t,k;m_{1,2})$ are given by equation (\ref{I1f}) for $m=m_{1,2}$ respectively. In terms of this total contribution  the decay of the quasiparticle is described by
 \be \alpha_k(t) = \frac{Y^2_1}{8\pi^2\,\gamma_k}\, I^{tot}_1(t,k)\,. \label{alfantizeno}\ee

\begin{figure}[ht!]
\begin{center}
\includegraphics[height=3in,width=3in,keepaspectratio=true]{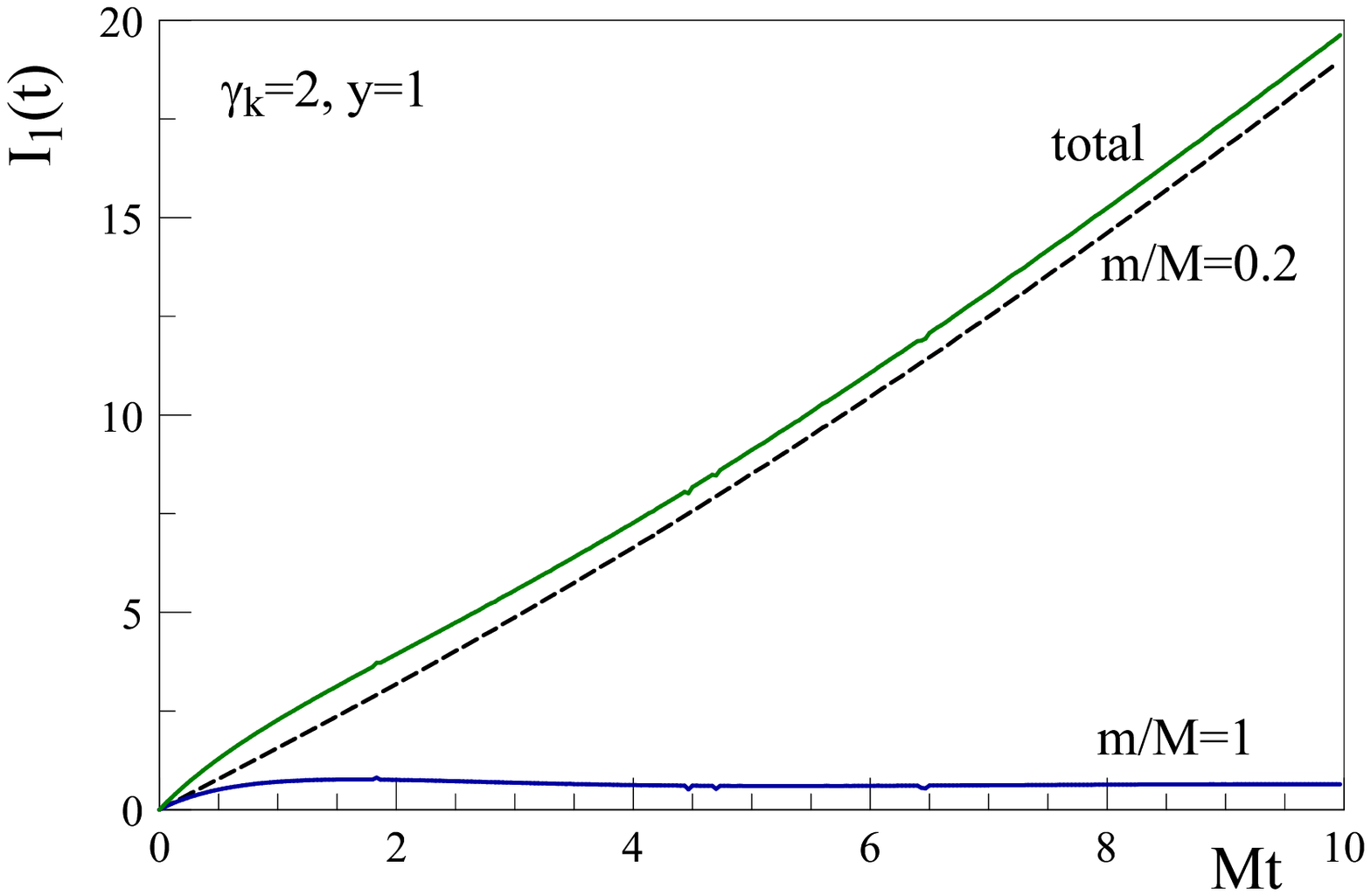}
\includegraphics[height=3in,width=3in,keepaspectratio=true]{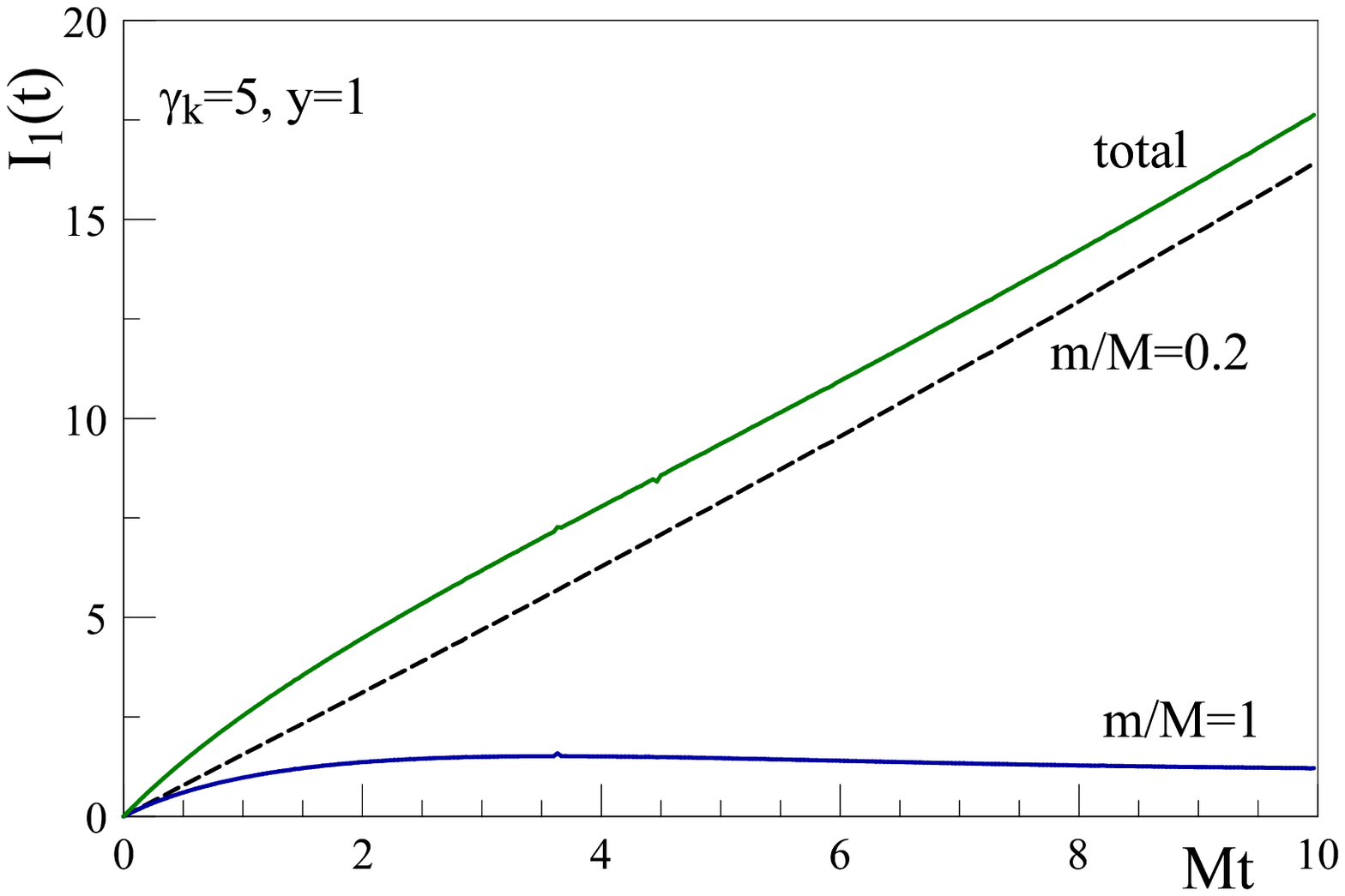}
\caption{Antizeno effect for the renormalizable case. $I_1(t)$ vs $Mt$ for $m/M=0.2$ and $m/M=2$ with $y=1$. The solid line ``\emph{total}'' is the sum for the total spectral density. The  lines labeled $m/M=0.2, m/M=1$ yield  the contributions from the lower and higher mass spectral densities respectively. Left figure corresponds to $\gamma_k=2$, right figure to $\gamma_k=10$.    }
\label{fig:antizenorengama2y1}
\end{center}
\end{figure}

\begin{figure}[ht!]
\begin{center}
\includegraphics[height=3in,width=3in,keepaspectratio=true]{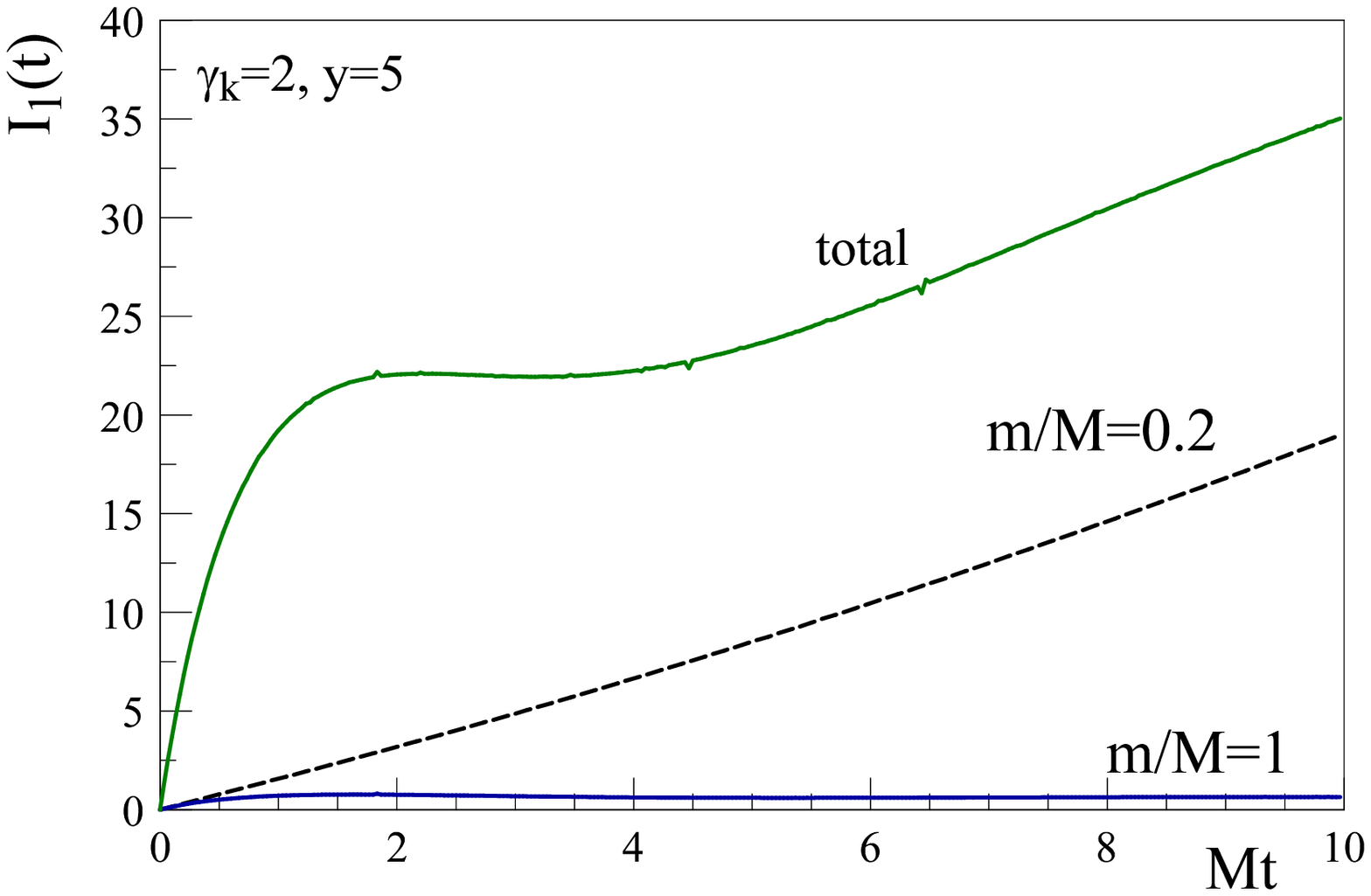}
\includegraphics[height=3in,width=3in,keepaspectratio=true]{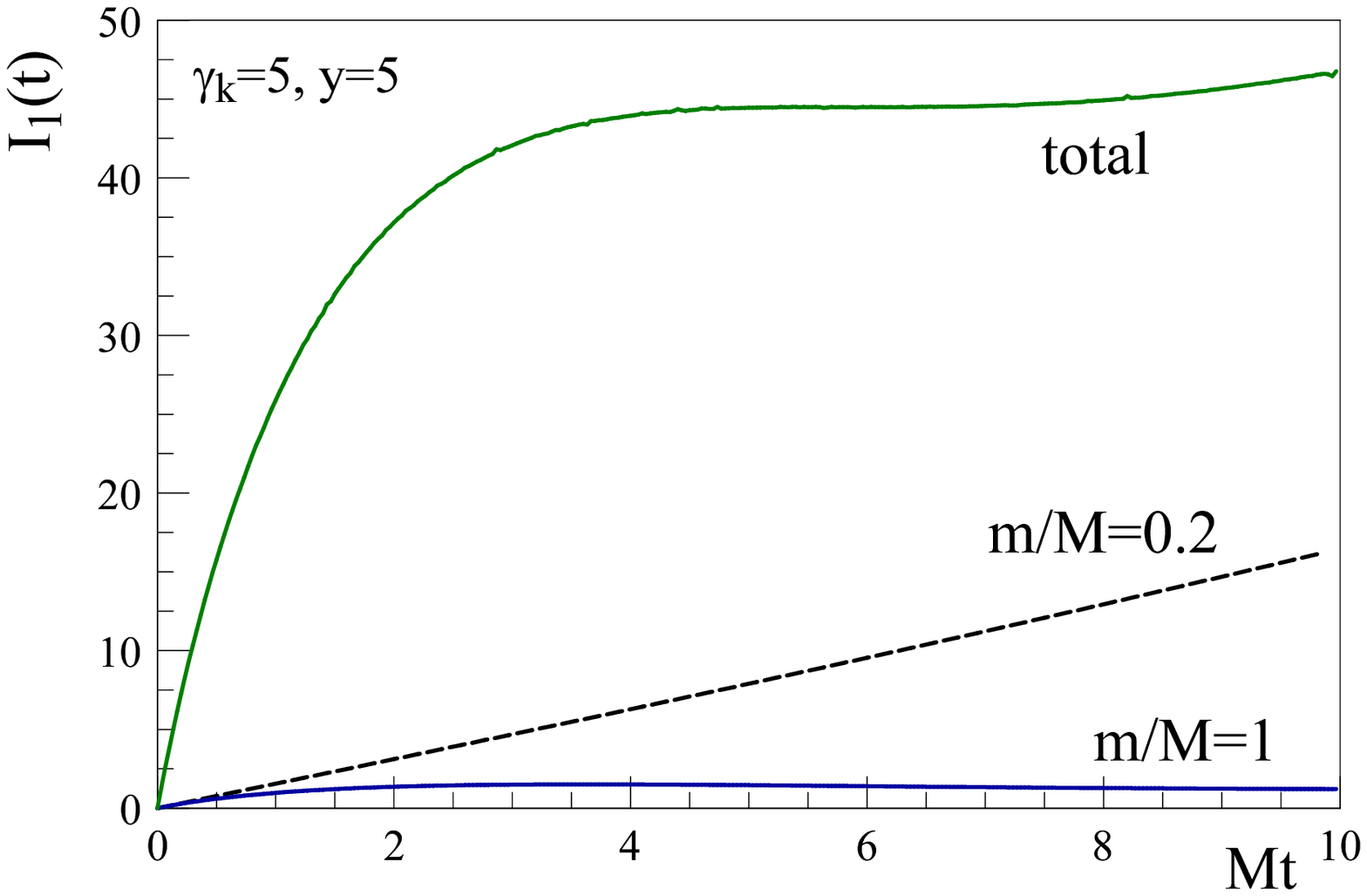}
\caption{Antizeno effect for the renormalizable case. $I_1(t)$ vs $Mt$ for $m/M=0.2$ and $m/M=1$ with $y=5$. The solid line ``\emph{total}'' is the sum for the total spectral density. The  lines labeled $m/M=0.2, m/M=1$ yield  the contributions from the lower and higher mass spectral densities respectively Left figure corresponds to $\gamma_k=2$, right figure to $\gamma_k=5$.    }
\label{fig:antizenorengama2y5}
\end{center}
\end{figure}

The numerical analysis confirms the main physical picture and the estimates on the time scale at which the higher mass channels closes given by (\ref{closing}). For $t \gg  t_c$  the contribution from the higher mass spectral density to the decay function levels off saturating at a constant value. This asymptotic limit is found by the following analysis of the integral $I_1(t,k;m_2)$ given by (\ref{I1f}):   for $m=m_2 > M/2$ the region of integration \emph{does not} include the origin therefore for $Mt \gg \Bigg[\Big(\gamma^2_k + \frac{4m^2_2}{M^2}- 1\Big)^{1/2}-\gamma_k   \Bigg]^{-1}$ the maximum of the function $(1-\cos(x))/x^2$ is well outside the domain of integration, hence  i) define $x= w M t$,   ii) now because the integral is finite  the   oscillatory term $\cos(w M t)$ in the integrand in (\ref{I1f})  yields  an oscillatory function with an amplitude that falls-off in time with a power law  by the Riemann-Lebesgue lemma (this is  confirmed numerically) and the contribution of the cosine term averages to zero at long time. Taking $\Lambda/M \rightarrow \infty$,  the asymptotic long time limit for $2m_2/M >1$ is given by
\be I_1(\infty,k;m_2) = \int^{\infty}_{W_{th}} \Bigg[1-\frac{4m^2}{M^2\,g(w)}\Bigg]^{3/2}\, \frac{dw}{w^2} \,,  \label{I1asy} \ee where
\be W_{th}= \Bigg[\Big(\gamma^2_k + \frac{4m^2_2}{M^2}- 1\Big)^{1/2}-\gamma_k   \Bigg]~~;~~ g(w) = 1+ 2\, \gamma_k\,w+w^2 \,.\label{funI1asy} \ee

\begin{figure}[ht!]
\begin{center}
\includegraphics[height=4in,width=4in,keepaspectratio=true]{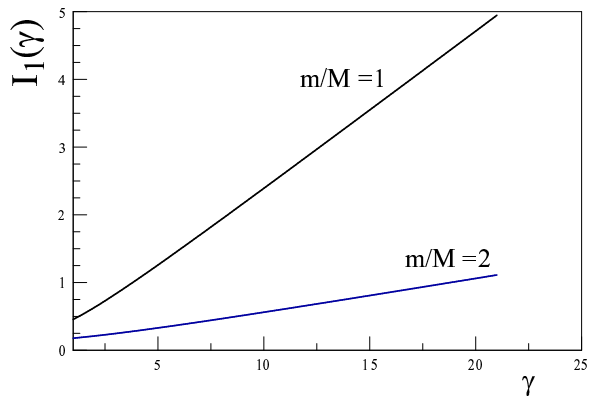}
\caption{The function $I_1(\infty,k;m) \equiv I_1(\gamma)$ vs. $\gamma$,  for $m_2/M=1,2$ respectively.    }
\label{fig:i1asy24}
\end{center}
\end{figure}

The total decay function asymptotically approaches a linear behavior with $t$ with a constant off-set $y^2\,I_1(\infty,k,m_2)$ from the accumulation of the contribution from the higher mass threshold. This finite asymptotic value \emph{cannot} be subtracted  away into a redefinition of the  quasiparticle   because i) the subtraction of the quasiparticle contribution at $t_b$ must vanish at $t_b$ and ii) it is the result of a time evolution on the \emph{slow} time scales, and for $\gamma_k \gg 1$ the asymptotic limit emerges at very long time.

The asymptotic function $I_1(\infty,k;m_2)\equiv I_1(\gamma)$ is shown in fig. (\ref{fig:i1asy24}) for $m_2/M=1,2$ respectively and confirms the asymptotics of the figures  (\ref{fig:antizenorengama2y5}). The main features of fig. (\ref{fig:i1asy24}) are clear: i) the growth of $I_1(\infty,k;m_2)$ with $\gamma_k$ is precisely the signature of the anti-Zeno effect: the lower limit $W_{th}$ in (\ref{funI1asy}) is precisely the distance between the ``pole'' at $\omega_k = M \gamma_k$ and the threshold $\sqrt{\omega^2_k+4m^2_2-M^2}$ in units of $M$. For large $\gamma_k$ this distance is reduced and the overlap with the time dependent function $C(k_0,t)$ is larger, consequently yielding the growth with $\gamma_k$. ii) The suppression with larger $m/M$ is also expected since the higher mass threshold is farther away from the pole. Since the contribution from the higher mass threshold is multiplied by the ratio of Yukawa couplings, in the standard model where fermions acquire masses via the expectation value of the Higgs field, the larger mass thresholds receive proportionately a larger contribution.

Depending on the values of Yukawa couplings the   time scale at which the usual decay function linear in time (up to a constant off-set) emerges could be quite long and a substantial decrease of the survival probability \emph{beyond that predicted by exponential decay} could result during this time interval. Therefore the asymptotic off-set of the decay function is a ``lingering'' memory of the anti-Zeno transient dynamics.

\section{Discussion}

\textbf{Real time evolution vs. S-matrix:}
The full time dependence of the decay function $\alpha_k(t)$, including transient dynamics,  cannot be obtained within the S-matrix formulation of a decay rate. In the latter one considers the transition amplitude from a single particle initial state prepared at time $t \rightarrow -\infty$, the ``in'' state,   to the ``out'' state with two bosons or fermions as $t\rightarrow +\infty$. In taking the infinite time limit this transition amplitude features an overall energy conservation delta function, upon squaring it to obtain the transition probability one extracts the total interaction time. Dividing by this total time defines the transition probability per unit time or decay rate $\Gamma_k$. The infinite time limit is equivalent to Fermi's Golden rule. Obviously, in taking the infinite time limit all the transient dynamics has been neglected and only the term growing linearly with total time is obtained yielding the usual exponential decay law. Although this is the standard procedure to obtain the decay rate it suffers from a conceptual caveat: a decaying state is prepared at $t\rightarrow -\infty$ and its transition amplitude is obtained as $t\rightarrow \infty$ long after it has decayed. Instead, the real time evolution within the non-perturbative Weisskopf-Wigner framework implemented above allows to study the transient dynamics of the formation of the quasiparticle state along with the details of the decay process. In particular the acceleration of the decay as a consequence of the energy uncertainty at finite time and   the proximity of the ``on-shell'' pole to a higher mass threshold cannot possibly be addressed within the S-matrix formulation, since in the infinite time limit the energy uncertainty vanishes, enforcing strict energy conservation.

There is another important technical difference between the S-matrix approach and the real time evolution, which is set up as an \emph{initial value problem} for a quantum state. The calculation of the one-loop self energy in quantum field theory in four-momentum space, features a quadratic divergence with a (Euclidean rotational invariant cutoff) which is absorbed into a mass renormalization, and a logarithmic divergence which also contributes to wave function renormalization, but does \emph{not} feature a linear ultraviolet divergence with an energy cutoff $\Lambda$. This is a consequence of a manifestly Lorentz (or Euclidean) invariant regularization.  However, it is a straightforward exercise to see that \emph{if} the loop integrals are cutoff in energy and spatial momentum and not just in the magnitude of the four momentum, there emerges a \emph{linear} cutoff dependence proportional to $\omega_k$. In dimensional regularization, a manifestly (Euclidean) invariant regularization,  the divergences are manifest as single poles in $D-4$ with $D$ being the space-time dimensionality. Because the real time evolution selects the time direction explicitly, the linear dependence in the energy cutoff becomes manifest and is unavoidable. Recently a thorough and deeper study of the Lehmann-Symanzik-Zimmerman reduction formula within a \emph{finite time analysis} also revealed a linear cutoff dependence of single particle state renormalization in a renormalizable theory\cite{collins}.

The difference in regularizations between the one-loop self energy and the real-time evaluation of the decay function notwithstanding, the renormalization procedure to absorb the short time behavior into the quasiparticle state at an intermediate time scale $t_b$ (namely the contribution from the functions $I_{1,2}$) yields a decay law valid for time scales much longer than that of the transient build up and which is insensitive to the details of the regularization prescription (cutoff).

\vspace{1mm}

\textbf{Transient dynamics vs. infinite time limit:} To be sure, in the limit $t\rightarrow \infty$ we find that the Lorentz boost suppression along with the anti-Zeno enhancement in the case of higher threshold channels saturate and the long time limit features a linear secular term in time describing exponential decay, in agreement with the S-matrix calculation and Fermi's Golden rule.   The transient dynamics saturates at different time scales determined by the Lorentz factors and the proximity to the higher mass thresholds. These time scales can be much longer than the ``natural'' scale $1/M$ substantially modifying the decay law during the transient stage.   The transient dynamics yields non-vanishing (and non-trivial) contributions to the decay function at long time, in the form of a constant off-set,  and the survival probability retains memory of the early stages in the form of either a   suppression from the anti-Zeno effect or enhancement because of the further delay for large Lorentz boosts. Therefore, we conclude that there are substantial caveats to the validity of the ``infinite time limit'' determined by the different processes during the transient stage, depending on  the Lorentz factors and the specific details of the higher mass thresholds and coupling strength.

As discussed above while we do not expect that transient phenomena will yield observable signatures in particle physics experiments, the Zeno suppression and anti-Zeno effects\cite{wineland, zenoobs} and quasiparticle formation\cite{birth} have been observed in condensed matter experiments. Therefore there are explicit examples of the physical manifestations of the transient effects discussed above.

\vspace{1mm}

\textbf{Implications for cosmology:}
Our study of the transient dynamics is ultimately motivated by understanding decay processes in cosmology where the time dependence associated with the expansion of the Universe implies the lack of particle energy conservation.  Describing decay (and in general quantum kinetic processes) via the S-matrix calculation of matrix elements is clearly an approximation that must be subject to a critical assessment. A recent study\cite{herring} of particle decay in \emph{superrenormalizable theory} in a radiation and matter dominated cosmology found non-trivial phenomena associated with cosmological expansion.  In particular  a similar phenomenon to the anti-Zeno acceleration has been reported  there  where the energy uncertainty is of the order of the Hubble rate of expansion. This phenomenon thus opens new decay channels into heavier particle states with their concomitant production  that will contribute to quantum kinetic processes that would be forbidden by strict energy conservation.  The dynamical aspects of the quasiparticle formation and renormalization effects studied in this article must now be incorporated into the framework developed in this reference to yield  a consistent   quantum kinetic description with the potential for novel processes of particle production and decay.

\section{Conclusions:}

The time evolution of unstable quantum states continues to be a subject of theoretical and experimental study with impact in various fields. Motivated by the ubiquity of the decay process in particle physics models and their fundamental importance in early Universe cosmology, in this article we   study the dynamics of  particle decay in \emph{renormalizable theories}. The usual quantum field theoretical approach to quantum decay, either by obtaining a decay rate from asymptotic S-matrix theory or via a Breit-Wigner approach to the propagators of unstable particles, are not suitable to describe particle decay in a rapidly expanding cosmology because of the lack of energy conservation and   of well defined asymptotic states.
Instead here we implement a quantum field theoretical extension of the Weisskopf-Wigner theory of atomic linewidth to obtain the explicit time evolution of an unstable state \emph{including} transient phenomena. The survival probability of an initially prepared quantum single particle state with momentum $\vk$ obeys a generic decay law $P_k(t) = e^{-\alpha_k(t)}\,P_k(0)$, in renormalizable theories the density of states grows rapidly with energy leading to ultraviolet divergences in   $\alpha_k(t)$. These divergences are associated with the renormalization of the bare state into a ``dressed'' renormalized quasiparticle state. Introducing a cutoff in energy $\Lambda$ we find that the transient dynamics that builds the quasiparticle state occurs on a time scale $1/\Lambda$, and this renormalized quasiparticle state decays on much longer time scales. During the stage in which the quasiparticle forms the decay law features a Zeno-like behavior  $\alpha_k(t)= (t/t_Z)^2$, where for a renormalizable quantum field theory $1/t^2_Z \propto \Lambda^3$. We introduce a consistent \emph{dynamical renormalization}  framework that exploits the wide separation of time scales between the formation of the quasiparticle and its decay, to separate both processes. This is achieved by introducing a sliding time scale $t_b$ at which the transient phenomena from quasiparticle formation has subsided, to absorb the ultraviolet divergences into a redefinition of the renormalized quasiparticle state. The total survival probability is independent of the choice of $t_b$ and obeys a renormalization group equation with respect to this (arbitrary) scale. We find a wealth of phenomena associated with transient dynamics that is \emph{not} captured by an S-matrix (``in-out'') approach: \textbf{i)} a delayed decay as a consequence of Lorentz boost different from the usual time dilation,    \textbf{ii)} an accelerated decay when there are higher mass thresholds resulting in an \emph{anti-Zeno} enhancement in the decay law. The delayed decay for highly boosted particles is a consequence of the narrowing of the phase space for large Lorentz factors. The anti-Zeno enhancement of the decay law is a consequence of the energy uncertainty associated with the early time dynamics, this effect is also enhanced by large Lorentz factor because of the diminishing distance between the position of the ``pole'' and the higher mass threshold for large boost. In the standard model where fermions acquire a mass from the expectation value of the Higgs field, more massive fermions feature larger Yukawa couplings, which in turn enhances the anti-Zeno contribution to the decay.

In the \emph{very long time limit} both effects, delay from Lorentz boost and acceleration from anti-Zeno effect, saturate and $\alpha_k(t) \propto t$. However the time scale for the exponential decay to dominate becomes very long for large Lorentz factor and when the mass of the decaying particle is close to (but below) the higher mass thresholds. This, in turn may yield a substantial \emph{production} of heavier states during the transient dynamics.

As discussed in ref.\cite{herring} the time dependence of the gravitational background in an expanding cosmology introduces an energy uncertainty $\propto H$ with $H$ the Hubble expansion rate, which may translate into a substantial anti-Zeno contribution to the decay and to the production of heavier states with potentially important implications in cosmology.    Thus we expect that while these transient phenomena are unlikely to bear direct impact in particle physics experiments, they are manifestations of effects that had been observed in condensed matter systems\cite{wineland,zenoobs,birth} and may prove to be very relevant in cosmology.

\acknowledgments DB gratefully acknowledges support from the US NSF through grant PHY-1506912. The author thanks John Collins for illuminating correspondence and comments.

\appendix
\section{Spectral densities}
The spectral density is defined by eqn. (\ref{specdens}), where the main ingredient is the matrix element $\bra{\kappa}H_i\ket{1^{\Phi}_k}$. In the superrenormalizable case the state $\ket{\kappa}$ corresponds to a bosonic pair while in the renormalizable case it is a fermion-anti-fermion pair.
\subsection{Superrenormalizable case:} With the Lagrangian density (\ref{Lcub}) and field quantization (\ref{boseexp}) we find
\be |\bra{1^{\chi}_{\vp};1^{\chi}_{\vq}}H_i\ket{1^{\Phi}_k}|^2 = \frac{4\,\lambda^2}{V^2\,2\omega_k \,2E_p\,2E_q} \,(2\pi)^3\,\delta^{3}(\vk-\vp-\vq) \ee and with
\be \sum_{\kappa} \rightarrow \frac{1}{2!}\sum_{\vp}\,\sum_{\vq} \rightarrow \frac{1}{2!}\,V\,\int\frac{d^3p}{(2\pi)^3}\,V\,\int\frac{d^3p}{(2\pi)^3} \label{sumis}\ee with $\varepsilon_\kappa = E_p+E_q$ the spectral density (\ref{specdens}) becomes
\be \rho(k_0,k) = \frac{\lambda^2}{ \omega_k}\,\int\frac{d^3p}{(2\pi)^3\,2E_p}\, \int\frac{d^3q}{(2\pi)^3\,2E_q} (2\pi)^3\,\delta(k_0-E_p-E_q)\,\delta^{3}(\vk-\vp-\vq) \,. \label{rhosupa}\ee The integrals are the familiar Lorentz invariant phase space for two body decay, leading to the result (\ref{rhoS}).

\subsection{Renormalizable:} With the Lagrangian density (\ref{Lyuk}) and fermion field quantization (\ref{fermiexp}), and the intermediate state with a fermion-antifermion pair of momenta $\vp,\vq$ respectively, we find
\be \bra{1_{\vp,s},\overline{1}_{\vq,s'}}H_i \ket{1^{\Phi}_k} = \frac{Y}{V^{3/2}}\sum_{a}\overline{U}_{\vp,s,a}\,V_{\vq,s',a} \frac{(2\pi)^3\,\delta^{3}(\vk-\vp-\vq) }{\Big[2\omega_k \,2E_p\,2E_q \Big]^{1/2}} \,. \label{renmtxel}\ee In this case
\be \sum_{\kappa} = \sum_{s}\sum_{s'}\sum_{\vp}\sum_{\vq} \rightarrow \sum_{s}\sum_{s'} V\,\int\frac{d^3p}{(2\pi)^3}\,V\,\int\frac{d^3p}{(2\pi)^3}\,, \label{sumkapfer}\ee using the identities (\ref{projec}) we   find
\be \rho(k_0,k) = \frac{2\,Y^2}{\omega_k}\,\int\frac{d^3p}{(2\pi)^3\,2E_p}\, \int\frac{d^3q}{(2\pi)^3\,2E_q} \Big[E_p E_q - \vp\cdot\vq - m^2 \Big] (2\pi)^3\,\delta(k_0-E_p-E_q)\,\delta^{3}(\vk-\vp-\vq)\,. \label{rhoyuki}\ee Using $k_0 = (E_p+E_q)$ and $\vq=\vk-\vp$, the bracket in (\ref{rhoyuki}) becomes
$\frac{1}{2} (k^2_0-k^2)\,\Big[ 1- \frac{4m^2}{(k^2_0-k^2)} \Big]$ and can be taken out of the integral. The remaining integral is the same as that for the two-body phase space (\ref{rhosupa}), yielding the result (\ref{rhoR}).

\end{document}